\documentclass[aps,prc,superscriptaddress,showpacs,nofootinbib,floatfix,amssymb,amsfonts,amsmath]{revtex4-1}
\usepackage{graphicx}
\usepackage{dcolumn}
\usepackage{bm}

\usepackage{amsmath}    
\usepackage{amsfonts}   
\usepackage{amssymb}
\usepackage{graphicx}   
\begin{document}

\title{Rotating nuclei: from ground state to the extremes of spin
                          and deformation \label{ch8}}

\author{A.\ V.\ Afanasjev}
\affiliation{Department of Physics and Astronomy, Mississippi
State University, MS 39762}

\date{\today}

\begin{abstract}
 The rotating nuclei represent one of most interesting subjects for theoretical
and experimental studies. They open a new dimension of nuclear landscape, 
namely, spin direction. Contrary to the majority of nuclear systems, their 
properties sensitively depend on time-odd mean fields and currents in 
density functional theories. Moreover, they show a considerable interplay of 
collective and single-particle degrees of freedom. In this chapter, I discuss 
the basic features of the description of rotating nuclei in one-dimensional 
cranking approximation of covariant density functional theory. The successes 
of this approach to the description of rotating nuclei at low spin in pairing 
regime and at high spin in unpaired regime in wide range of deformations (from 
normal to hyperdeformation) are illustrated. I also discuss the recent progress 
and open questions in our understanding of the role of proton-neutron pairing 
in rotating nuclei at $N\approx Z$, the physics of band termination and other
phenomena in rotating nuclei.
\end{abstract}



\maketitle
\tableofcontents


\section{Introduction}

  The development of self-consistent many-body theories aiming at the description 
of low-energy nuclear phenomena provides the necessary theoretical tools for an 
exploration of the nuclear chart into known and unknown regions. Theoretical 
methods (both relativistic \cite{Vretenar2005Phys.Rep.101} and non-relativistic 
\cite{Bender2003Rev.Mod.Phys.121}) formulated 
within the framework of density functional theory (DFT) are the most promising 
tools for the global investigation of the properties of atomic nuclei. The power 
of the DFT models is essentially unchallenged in medium and heavy mass nuclei where 
'ab-initio' type few-body calculations are computationally impossible and the 
applicability of spherical shell model is restricted to a few regions in the 
vicinity of doubly shell closures.

  One can consider a nuclear chart as three-dimensional object in which the 
charge, isospin and spin play a role of the coordinates. The studies  of superheavy 
and  neutron-rich nuclei allow to extend this chart in the directions of the charge 
and isospin, respectively. On the other hand, the investigations of rotating 
nuclei explore nuclear chart in the spin direction. Over the decades, the studies 
of rotating nuclei have revealed a lot of new interesting nuclear
phenomena such as superdeformation \cite{BHN.95}, smooth band termination 
\cite{Afanasjev1999Phys.Rep.1},  magnetic rotation \cite{Frauendorf2001Rev.Mod.Phys.463} 
etc. Many of them can be succesfully studied in the framework of one-dimensional 
cranking approximation \footnote{The review of two- and three-dimensional cranking 
approximations in the CDFT is presented in Ref.\ \cite{Meng2013Front.Phys.55} and 
in Chapter 9 of this book.}. The realization of this approximation in the framework 
of covariant density functional theory (CDFT) and its application to rotating nuclei 
in different regimes of spin, deformation and pairing and in different regions of 
nuclear chart are reviewed in the present paper. This review mostly focuses on the 
physical phenomena which were studied from 2005 since pre-2005 studies of rotating 
nuclei in the CDFT have been overviewed in Ref.\ \cite{Vretenar2005Phys.Rep.101}.  However, few examples 
of the studies of rotating nuclei from this time period are given in Sect.\ 
\ref{other-phenomena} for a completeness of the picture.

\section{One-dimensional cranking approximation}

 The description of rotating nuclei requires a transformation of the 
relativistic Hartree-Bogoliubov (RHB) equations to the rotating frame. 
The concept of a frame, which rotates around a fixed axis with constant 
rotational frequency $\Omega$, is not strictly compatible with Lorentz 
invariance because a rotating frame is an accelerated rather than an inertial one.
Strictly speaking such a problem should be treated in the framework
of general relativity. However, one can also look on this problem
differently. Points connected with such a frame move with the
velocity $r\Omega$, where $r$ is the distance of such a point from
the axis of rotation. In nuclei the relevant distances $r$ are not
much larger than the nuclear radius $R$. Since rotational
frequencies in nuclei usually do not exceed $\Omega \sim 1-2$ MeV,
the linear velocities at the surface of rotating nuclei are of the
order of only a few percents of the velocity of light. Therefore, the
effects of general relativity can be neglected and we just have to
transform the system to a coordinate frame rotating with constant
rotational frequency $\Omega$. Such transformations have been
studied in detail within the semi-classical approximation
\cite{Koe.96}, within the formalism of special relativity
\cite{Koepf1989Nucl.Phys.A61} and with techniques of general relativity 
\cite{Madokoro1997Phys.Rev.C2934}.
All these investigations lead to identical results.

  The transformation of the RHB equations to the rotating frame leads
to the cranked RHB (CRHB) equations \cite{Afanasjev1999Phys.Rev.C51303,Afanasjev2000Nucl.Phys.A196}. 
Note that in 
the present manuscript only one-dimensional rotation with rotational 
frequency $\Omega_x$ around $x$-axis ({\it one-dimensional cranking 
approximation}) is considered. Their nucleonic part has the form
\begin{equation}
\left(
\begin{array}{cc}
\hat{h}_{D}-\lambda _{\tau }-\Omega_x\hat{j_x} & \hat{\Delta}
\\
-\hat{\Delta}^{\ast } & -\hat{h}_{D}^{\ast }+\lambda _{\tau }+ \Omega_x \hat{j_x}^{\ast }%
\end{array}%
\right) \left(
\begin{array}{l}
U({\bf r}) \\
V({\bf r})%
\end{array}%
\right) _{k}= E_{k} \left(
\begin{array}{l}
U({\bf r}) \\
V({\bf r})%
\end{array}%
\right) _{k}  
\label{eq.ch8:1}
\end{equation}
with the chemical  potentials $\lambda _{\tau }$ ($\tau =n,p)$ for
neutrons and protons and the single-particle angular momentum 
operators $\hat{j_x}$ for 
fermions with spin $\frac{1}{2}$.
%
%
The Dirac  Hamiltonian $\hat{h}_{D}$
\begin{equation}
\hat{h}_{D}=\bm{\alpha}(-i\bm{\nabla}-\bm{V}({\bf r}))~+~V_{0}({\bf
r})~+~\beta (m+S({\bf r}))
\label{eq.ch8:2}
\end{equation}
contains the average fields determined by the mesons, i.e. the attractive 
scalar field $S({\bf r})$
\begin{eqnarray}
S({\bf r})=g_{\sigma} \sigma ({\bf r}),
\label{eq.ch8:3}
\end{eqnarray}
and the repulsive time-like component of the vector field $V_{0}({\bf r})$
\begin{eqnarray}
V_0({\bf r}) = g_{\omega} \omega_0({\bf r}) + g_{\rho} \tau_3 \rho_0 ({\bf r}) + e 
\frac{1-\tau_3}{2} A_0 ({\bf r}).
\label{eq.ch8:4}
\end{eqnarray}  
 A magnetic potential ${\bm V} ({\bf r})$ 
\begin{equation}
\bm{V}({\bf r})=g_{\omega }\bm{\omega}({\bf r})+ g_{\rho
}\tau_3\bm{\rho}({\bf r})+e\frac{1-\tau _{3}}{2}\bm{A}({\bf r}),
\label{eq.ch8:5}
\end{equation}
originates from the space-like components of the vector mesons and
behaves in the Dirac equation like a magnetic field. Therefore
the effect produced by it is called \textit{nuclear magnetism } 
\cite{Koepf1989Nucl.Phys.A61}.

   Note that in these equations, the four-vector components of the vector 
fields $\omega^{\mu}$, $\rho^{\mu}$, and $A^{\mu}$ are separated into the time-like 
($\omega_0$, $\rho_0$ and $A_0$) and space-like 
[${\bm \omega}=(\omega^x, \omega^y, \omega^z)$, 
${\bm \rho}=(\rho^x, \rho^y, \rho^z)$, and ${\bm A}=(A^x, A^y, A^z)$] components. 

The cranked RHB-equations (\ref{eq.ch8:1}) contain three constraints
characterized by three Lagrange parameters, the chemical potentials
$\lambda _{\tau }\;(\tau =n,p)$ and the rotational frequency
$\Omega_x$. The chemical potentials $\lambda _{\tau }$ are determined
by the average particle numbers for neutrons and protons
($\tau=n,p$)
\begin{equation}
\qquad \langle {\Phi }_{\Omega }|\hat{N}_{n}|{\Phi }_{\Omega
}\rangle =N,\;\;\;\;\ \langle {\Phi }_{\Omega }|\hat{N}_{p}|{\Phi
}_{\Omega }\rangle =Z.
\label{eq.ch8:6}
\end{equation}
  The rotational frequency $\Omega_x $ along the $x$-axis is
defined from the condition \cite{Inglis1956Phys.Rev.1786}
\begin{equation}
J_x(\Omega_x )=\langle {\Phi }_{\Omega_x }\mid \hat{J}_{x}\mid {\Phi
}_{\Omega_x }\rangle=\sqrt{I(I+1)}.  
\label{eq.ch8:7}
\end{equation}%
where $J_x(\Omega_x)$ is the expectation value of the total angular momentum 
$\bm{\hat{J}}$ of the system along the $x$-axis and $I$ is the total 
nuclear spin. The angular momentum is carried essentially by the 
fermions. As a result, the contributions of the meson fields to 
$J_x(\Omega_x)$ are neglected.
The Coriolis term is given by 
\begin{eqnarray}
-\Omega_x  \hat{J}_x = -\Omega_x \left( \hat{L}_x+\frac{1}{2}\hat{\Sigma}_x \right). 
\label{eq.ch8:8}
\end{eqnarray}

  The time-independent inhomogeneous Klein-Gordon equations for the 
mesonic fields obtained by means of variational principle are given 
in the CRHB theory by \cite{Afanasjev1999Phys.Rev.C51303,Afanasjev2000Nucl.Phys.A196}
\begin{eqnarray}
\left\{-\Delta-({\Omega}_x\hat{L}_x)^2 + m_\sigma^2\right\}~
\sigma(\bf r) & = & 
-g_\sigma \rho_{\sl s}({\bf r}) - g_2\sigma^2({\bf r})-g_3\sigma^3({\bf r}),
\nonumber \\
\left\{-\Delta-({\Omega}_x\hat{L}_x)^2+m_\omega^2\right\}
\omega_0({\bf r})&=&
g_\omega \rho_{\sl v}^{\it i\sl s}({\bf r}),
\nonumber \\
\left\{-\Delta-[{\Omega}_x(\hat{L}_x+\hat{S}_x)]^2+
m_\omega^2\right\}~
{\bm \omega}({\bf r})&=&
g_\omega{\bf j}^{\it i\sl s}({\bf r}),
\nonumber \\
\left\{-\Delta-({\Omega}_x\hat{L}_x)^2+m_\rho^2\right\}
\rho_0({\bf r})&=&
g_\rho\rho_{\sl v}^{\it i\sl v}({\bf r}),
\nonumber \\
\left\{-\Delta-[{\Omega}_x(\hat{L}_x+\hat{S}_x)]^2+
m_\rho^2\right\}~
{\bm \rho}({\bf r})&=& g_\rho {\bf j}^{\it i\sl v}({\bf r}),
\nonumber \\
-\Delta~A_0({\bf r})=e\rho_{\sl v}^p({\bf r}),\quad \quad-\Delta~{\bf A}({\bf r})&=&e{\bf j}^p({\bf r}),
\label{eq.ch8:9}
\end{eqnarray}
where the source terms are sums of bilinear products of 
baryon amplitudes
\begin{eqnarray}
\rho_{\sl s}({\bf r}) & = & \sum_{k>0} 
 [V_k^n({\bf r})]^{\dagger} \hat{\beta} V_k^n ({\bf r}) 
+[V_k^p({\bf r})]^{\dagger} \hat{\beta} V_k^p ({\bf r}), 
\nonumber \\
\rho_{\sl v}^{\it i\sl s}({\bf r}) & = & \sum_{k>0} 
 [V_k^n({\bf r})]^{\dagger} V_k^n ({\bf r}) 
+[V_k^p({\bf r})]^{\dagger} V_k^p ({\bf r}), 
\nonumber \\
\rho_{\sl v}^{\it i\sl v}({\bf r}) & = & \sum_{k>0} 
[V_k^n({\bf r})]^{\dagger} V_k^n ({\bf r}) 
-[V_k^p({\bf r})]^{\dagger} V_k^p ({\bf r}), 
\nonumber \\
\bf j^{\it i\sl s}({\bf r}) & = & \sum_{k>0} 
[V_k^n({\bf r})]^{\dagger} \hat{\bm\alpha} V_k^n ({\bf r}) 
+[V_k^p({\bf r})]^{\dagger} \hat{\bm\alpha} V_k^p ({\bf r}), 
\nonumber \\
\bf j^{\it i\sl v}({\bf r}) & = & \sum_{k>0} 
[V_k^n({\bf r})]^{\dagger} \hat{\bm\alpha} V_k^n ({\bf r}) 
-[V_k^p({\bf r})]^{\dagger} \hat{\bm\alpha} V_k^p ({\bf r}). 
\label{eq.ch8:10}
\end{eqnarray}
 The sums over $k>0$ run over all quasiparticle states corresponding
to positive energy single-particle states ({\it no-sea approximation})
\cite{Serot1986Adv.Nucl.Phys.1,NL1}. 
In Eqs.\ (\ref{eq.ch8:9},\ref{eq.ch8:10}), the indexes $n$ and $p$ indicate 
neutron and proton states, respectively, and the indexes $\it i\sl s$ 
and $\it i\sl v$ are used for isoscalar and isovector quantities.
$\rho_{\sl v}^p(\bf r)$, $\bf j^p(\bf r)$ in Eq.\ (\ref{eq.ch8:9})
correspond to $\rho_{\sl v}^{\it i\sl s}(\bf r)$ and 
$\bf j^{\it i\sl s}(\bf r)$ defined in Eq.\ (\ref{eq.ch8:10}), 
respectively, but with the sums over neutron states neglected. 
$\bm{\hat{L}}=-i{\bf r}\times\bm{\nabla}$ is the orbital angular 
momentum operator of the scalar and the time-like meson fields and 
$\bm{\hat{L}}+\bm{\hat{S}}$ is the total angular momentum 
of the space-like parts of the vector fields. The operators
$\mathbf{\hat{S}}$ are three 3$\times$3 matrices representing the
spin-matrices of vector fields (see \cite{Edmonds1957}). Note that
only $x$-components of these operators are considered in Eq.\ 
(\ref{eq.ch8:10}). The Coriolis terms $\Omega_x (\hat{L}_x+\hat{S}_x)$ 
and $\Omega_x \hat{L}_x$ can be neglected for reasons discussed in 
Sect.\ 8.1 of Ref.\ \cite{Vretenar2005Phys.Rep.101}. The Coriolis term for the Coulomb 
potential $A_0({\bf r})$ and the spatial components of the vector potential 
$\bf A(\bf r)$ are neglected in Eqs.\ (\ref{eq.ch8:9}) since the coupling 
constant of the electromagnetic interaction is small compared with the 
coupling constants of the meson fields.

\begin{figure}
\centering
\includegraphics[width=13.0cm]{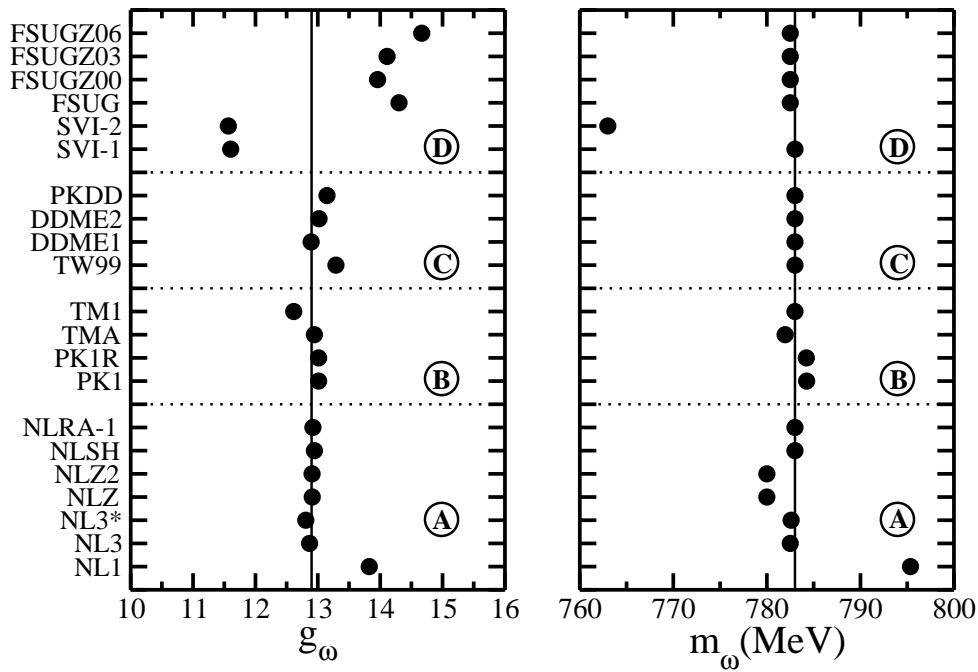}
\caption{The $m_{\omega}$ and $g_{\omega}$ parameters of different CEDF's. These CEDF's 
are combined into four groups dependent on how self- and mixed-couplings are 
introduced.  Group A represents the CEDF's which include non-linear self-couplings 
only for the $\sigma$-meson.  Group B contains the CEDF's which include self-couplings 
for the $\sigma$- and $\omega$-mesons (and $\rho-$mesons in the case of PK1R). Group 
C represents the CEDF's which include density-dependent meson-nucleon couplings for 
the $\sigma$-, $\omega$-, and  $\rho$-mesons. The other CEDF's are included into the 
group D. For details see Ref.\ \cite{Afanasjev2010Phys.Rev.C14309} from which this figure is taken.}
\label{fig.ch8:1}
\end{figure}

  Two terms in the Dirac equation, namely, the Coriolis operator $\hat{J}_x$ 
and the magnetic potential $\bf{V}({\bf r})$ (as well as the currents 
${\bf j}^{n,p}({\bf r})$ in the Klein-Gordon equations) break time-reversal 
symmetry \cite{AR.00}. Their presence leads to the appearance of time-odd mean 
fields. However, one should distinguish time-odd mean fields originating
from Coriolis operator and magnetic potential. The Coriolis operator is always
present in the description of rotating nuclei in the framework of the cranking
model. However, the cranked relativistic mean field (CRMF) calculations, with 
only these time-odd fields accounted for, underestimate the experimental moments 
of inertia \cite{Konig1993Phys.Rev.Lett.3079,AR.00,Afanasjev2010Phys.Rev.C34329}.
A similar situation also holds in nonrelativistic theories \cite{DD.95,YM.00}.
The inclusion of the currents ${\bf j}^{n,p}({\bf r})$ into the Klein-Gordon
equations, which leads to the space-like components of the vector $\omega$ and 
$\rho$ mesons and thus to magnetic potential ${\bf V}({\bf r})$, considerably 
improves the description of experimental moments of inertia.

 Note that time-odd mean fields related to nuclear magnetism are defined through 
the Lorentz invariance 
\cite{Vretenar2005Phys.Rep.101,Afanasjev2010Phys.Rev.C14309,Afanasjev2010Phys.Rev.C34329} 
and thus they do not require 
additional coupling constants: the coupling constants of time-even mean fields 
are used  also for time-odd mean fields. This is because the densities $\rho$ and 
currents $\bm{j}$ (Eq.\ (\ref{eq.ch8:10})) do not transform independently under 
Lorentz transformation since they form covariant four-vector $j^{\mu}=\{\rho, \bf{j}\}$. 
This fact explains why the structure of the Klein-Gordon equations for time-like and 
space-like components of vector mesons is the same (Eq.\ (\ref{eq.ch8:10})) and why the same 
coupling constant stands in front of the densities and currents on the right hand 
side of these equations \cite{Afanasjev2010Phys.Rev.C14309}.

  The currents are isoscalar and isovector in nature for the $\omega$ and $\rho$ 
mesons (Eqs.\ (\ref{eq.ch8:9} and \ref{eq.ch8:10})), respectively. As a consequence, the 
contribution of the $\rho$-meson to magnetic potential is marginal \cite{Afanasjev2010Phys.Rev.C14309}.
Thus, time-odd mean fields in the CDFT framework depend predominantly on the 
spatial components of the $\omega$ meson. Neglecting the contribution of the
$\rho$ meson, one can see that only two parameters, namely, the mass $m_{\omega}$  
and coupling constant $g_{\omega}$ of the $\omega$ meson define the properties of 
time-odd mean fields (Eqs.\ (\ref{eq.ch8:5}) and (\ref{eq.ch8:9})). 
Fig.\ \ref{fig.ch8:1} clearly indicates that these parameters are well localized 
in the parameter space for the covariant energy density functionals (CEDF's)  
in the groups A, B, and C (Fig.\ \ref{fig.ch8:1}). 
These are precisely the CEDF's which are extensively tested on nuclear structure 
data sensitive to time-even mean fields. For these CEDF's parameter dependence of 
the impact of time-odd  mean fields on the physical observables is quite weak
\cite{A.08,Afanasjev2010Phys.Rev.C14309,Afanasjev2010Phys.Rev.C34329}. This 
parameter dependence is expected to be larger in 
group D. However, at present it is not clear whether the CEDF's in this group
are reasonable ones since they have not been tested on nuclear structure data.

  Time-odd mean fields appear only in nuclear systems with broken time-reversal 
symmetry. They affect  magnetic moments \cite{Hofmann1988Phys.Lett.B307}, isoscalar monopole vibrations 
\cite{Engel1975Nucl.Phys.A215}, electric giant resonances \cite{NKKVR.08}, large amplitude collective 
dynamics \cite{HNMM.06}, fussion process \cite{UO.06}, the strengths and energies of 
Gamow-Teller resonances \cite{BDEN.02}, the binding energies of odd-mass nuclei 
\cite{S.99,DBHM.01,A250} and the definition of the strength of pairing correlations 
\cite{RBRM.99,A250,Afanasjev2010Phys.Rev.C14309}. However, 
as discussed in detail in Sect.\ \ref{TO-rotating} 
they are especially  pronounced in rotating nuclei.

  The pairing potential (field) $\hat{\Delta}$ in Eq.\ (\ref{eq.ch8:1}) 
is given by
\begin{eqnarray}
\hat{\Delta} \equiv \Delta_{ab}~=~\frac{1}{2}\sum_{cd} V^{pp}_{abcd} 
\kappa_{cd}
\label{eq.ch8:11}
\end{eqnarray}
where the indices $a,b,\dots$ denote quantum numbers which specify 
the single-particle states with the space coordinates
$\bf r$ as well as the Dirac and isospin indices $s$ and 
$\tau$. It contains the pairing tensor $\kappa$ \footnote{This quantity
is sometimes called as {\it abnormal density}.}
\begin{eqnarray}
\kappa = V^{*}U^{T} 
\label{eq.ch8:12}
\end{eqnarray}
and the matrix elements $V^{pp}_{abcd}$ of the effective interaction in 
the $pp$-channel.  The phenomenological Gogny D1S finite range interaction 
\cite{D1S-a} is used as such effective interaction in all CRHB calculations. 
It is given by 
\begin{eqnarray}
V^{pp}(1,2) = f \sum_{i=1,2} e^{-[({\bf r}_1-{\bf r} _2)/\mu_i]^2} 
\times (W_i+B_i P^{\sigma}- H_i P^{\tau} - M_i P^{\sigma} P^{\tau})
\nonumber \\
\label{eq.ch8:13}
\end{eqnarray}
where $\mu_i$, $W_i$, $B_i$, $H_i$ and $M_i$ $(i=1,2)$ are the 
parameters of the force and $P^{\sigma}$ and $P^{\tau}$ are the
exchange operators for the spin and isospin variables, respectively. 
This interaction is density-independent. Note also that an additional 
factor $f$ affecting the strength of the Gogny force is introduced 
in Eq.\ (\ref{eq.ch8:13}) (see Refs.\ \cite{A250,WSDL.13,Agbemava2014Phys.Rev.C54320} for
the reasons of its introduction).

  In the CRHB calculations, the size of the pairing correlations is measured 
in terms of the pairing energy defined as 
\begin{eqnarray}
E_{pairing}~=~-\frac{1}{2}\mbox{Tr} (\Delta\kappa).
\label{eq.ch8:14}
\end{eqnarray}
Note that this is not an experimentally accessible quantity.

 One should note that the Bogoliubov transformation is not commutable 
with the nucleon number operator and consequently the resulting wave 
function does not correspond to a system having a definite number of 
protons and neutrons. The best way to deal with this problem 
would be to perform an exact particle number projection 
before the variation \cite{Ring1980}; however,  such calculations are 
expected to be extremely time-consuming for realistic interactions.
As a result, an approximate particle number projection by means of 
the Lipkin-Nogami (LN) method \cite{L.60,N.64,NZ.64,PNL.73} 
is used in the CRHB calculations because of its simplicity. The details 
of the implementation of this method into the CRHB framework are given in 
Ref.\ \cite{Afanasjev2000Nucl.Phys.A196}; the CRHB calculations with the LN method included
are abbreviated as CRHB+LN ones. The  application of the LN method 
considerably improves an agreement with experiment for rotational 
properties (see Refs.\ \cite{Afanasjev2000Nucl.Phys.A196,J1Rare,AO.13}).

 The total energy of system in the laboratory frame is given as 
a sum of fermionic ( $E^{\rm F}$) and bosonic  ($E^{\rm B}$) 
contributions
\begin{eqnarray}
E_{\rm CRHB} = E^{\rm F}+E^{\rm B}.
\label{eq.ch8:15}
\end{eqnarray}
The fermionic energies $E^{\rm F}$ are given by
\begin{eqnarray}
E^{\rm F}=E_{part}  + \Omega_x J + E_{pairing}+ E_{cm} 
\label{eq.ch8:16}
\end{eqnarray}
where
\begin{eqnarray}
E_{part}=Tr (h_D \rho),\qquad \qquad J=Tr (j_x \rho),
\label{eq.ch8:17}
\end{eqnarray} 
are the particle energy and the expectation value of the total 
angular momentum along the rotational axis and
$E_{cm}$ is the correction for the spurious center-of-mass motion. 

  The bosonic energies $E^{\rm B}$ in the laboratory frame are
given by
\begin{eqnarray}
E^{\rm B}= && -~\frac{1}{2}\int d {\bf r} \,[g_\sigma\,\sigma {(\bf r)} 
\rho_{\sl s} {(\bf r)} + \frac{1}{3}g_2\sigma^3({\bf r}) 
+\frac{1}{2}g_3\sigma^4(\bf r)]  
\nonumber\\
&&-~\frac {1}{2}\, g_\omega \int d {{\bf r}}\,[\omega_0({\bf r}) 
\rho_{\sl v}^{\it i\sl s} ({\bf r}) 
          -{\bm\omega} ({\bf r}) {\bf j}^{\it i\sl s}({\bf r})]
\nonumber\\
&&-~\frac{1}{2}\, g_\rho \int d {\bf r}\,[\rho_0({\bf r}) 
\rho_{\sl v}^{\it i\sl v} ({\bf r})
-{\bm\rho} ({\bf r}) {\bf j}^{\it i \sl v} ({\bf r})]  
\nonumber\\
&&-~\frac{1}{2}\, e \int d {\bf r} \left[A_0({\bf r}) \rho_{\sl v}^p ({\bf r}) 
+ {\bm A} ({\bf r}) {\bf j}^p ({\bf r}) \right]  
\nonumber\\
&&+~{\Omega}_x^2\int d {\bf r} \,
[\sigma({\bf r}) \hat{L}_x^2\sigma({\bf r})- 
\omega_0({\bf r}) \hat{L}_x^2\omega_0({\bf r}) 
+{\bm\omega}({\bf r}) (\hat{L}_x+\hat{S}_x)^2
{\bm\omega} ({\bf r})
\nonumber \\
&&\qquad\qquad\qquad
-\rho_0({\bf r})\hat{L}_x^2\rho_0({\bf r})
+{\bm\rho}({\bf r})(\hat{L}_x+\hat{S}_x)^2{\bm\rho}({\bf r})]
\label{eq.ch8:18}
\end{eqnarray}

  Cranked relativistic mean field (CRMF) theory is a limiting case of the CRHB 
theory in which the pairing correlations are neglected; the details of the
formalism can be found in Refs.\ \cite{Koepf1989Nucl.Phys.A61,Konig1993Phys.Rev.Lett.3079,AKR.96}. It has been an
important step in the development of the CDFT theory to the description
of rotating nuclei and, as exemplified in the present paper, still remains 
a powerfull theoretical tool for the studies of rotating nuclei in unpaired
regime at high spin.

\section{Time-odd mean fields in rotating nuclei}
\label{TO-rotating}

  Nuclear magnetism, i.e. the time-odd component of the mean fields
$\bm{V}(\bf r)$, appears only in nuclear systems with broken
time-reversal symmetry in the intrinsic frame.  Rotating nuclei 
represent a system which is strongly affected by time-odd mean 
fields. In rotating nuclei, the average field has two sources of
time-reversal symmetry breaking, the Coriolis operator
$\Omega\hat{J}_{x}$ and the magnetic part of the vector fields
$\bm{V}(\bf r)$(Eq.\ (\ref{eq.ch8:5})) induced in a self-consistent
way by the currents. In this section, the results obtained with
and without nuclear magnetism are denoted as NM and WNM, respectively.

\begin{figure}[ptb]
\begin{center}
\includegraphics[width=18cm,angle=0]{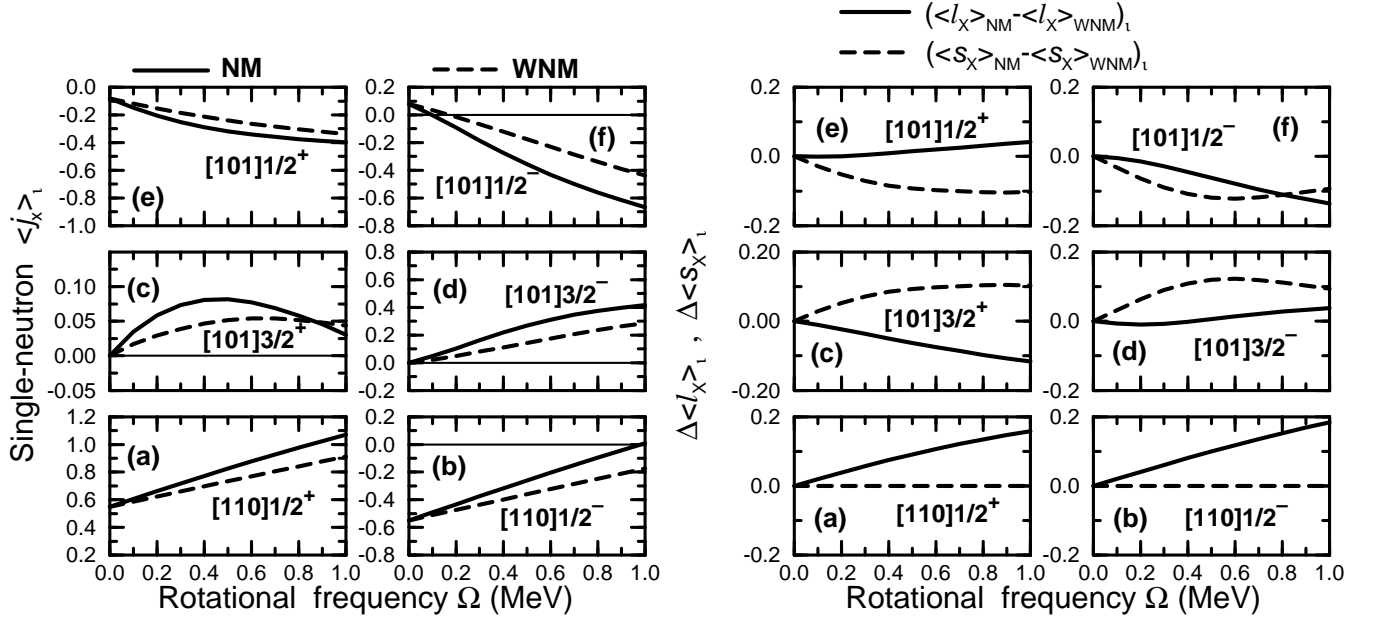}
\end{center}
\caption{Left two columns: expectation values $\langle{\hat j}_{x}
\rangle _{i}$ of the neutron orbitals forming the $N=1$ shell
calculated with and without nuclear magnetism. The orbitals are 
labeled by the Nilsson quantum numbers $[Nn_{z}\Lambda]\Omega$ and 
the sign of the signature $r=\pm i$. Right two columns: the changes 
of the expectation values of spin
and orbital angular momenta caused by nuclear magnetism. The results
presented are from the study of doubly magic superdeformed configuration 
in $^{152}$Dy. From Ref.\ \cite{AR.00}. }
\label{fig.ch8:2}
\end{figure}

  The physical observables, most frequently used in the analysis of rotating
nuclei, are kinematic ($J^{(1)}$) and dynamic ($J^{(2)}$) moments of inertia 
which are defined as
\begin{eqnarray}
J^{(1)}(\Omega_x)=\frac{J}{\Omega_x},\qquad 
J^{(2)}(\Omega_x)=\frac{dJ}{d\Omega_x}
\label{eq.ch8:19}
\end{eqnarray}
where $J$ is the expectation value of the total angular momentum along 
the $x$-axis. In the CRMF theory, this quantity is defined 
as a sum of the expectation values of the single-particle angular momentum 
operators $\hat{\jmath}_{x}$ of the occupied states
\begin{eqnarray}
J=\sum_{i}\langle i|\hat{\jmath}_{x}|i\rangle .
\label{eq.ch8:20}
\end{eqnarray}
Thus, the modifications of the moments of inertia due to NM, which as discussed
below are important in rotating nuclei, can be traced back to the changes of the 
single-particle expectation values 
$\langle \hat{\jmath}_{x}\rangle _{i}=$ $\langle i|\hat{\jmath}_{x}|i\rangle $ 
and the corresponding contributions of spin ($\langle \hat{s}_{x}\rangle _{i}$) 
and orbital ($\langle \hat{l}_{x}\rangle _{i}$) angular momenta \cite{AR.00}.

  On the microscopic level, the contribution to  $\langle \hat{j}_x \rangle_i$ 
due to NM is defined as \cite{AR.00}
\begin{eqnarray}
\Delta \langle j_x \rangle_i= \langle \hat{j}_x \rangle^{NM}_i
- \langle \hat{j}_x \rangle^{WNM}_i .
\label{eq.ch8:21}
\end{eqnarray}
Fig. \ref{fig.ch8:2} shows that the $\Delta \langle j_x \rangle_i$ 
is positive at the  bottom and negative at the top of the $N$-shell \cite{AR.00}. The 
absolute value of $\Delta \langle j_x \rangle_i$ correlates with the absolute value of 
$\langle \hat{j}_x \rangle_i$.  Note that the contributions to 
$\langle \hat{j}_x \rangle_i$ due to NM are small in the middle of the shell.
The $\Delta \langle j_x \rangle_i$ contributions can be decomposed into the contributions 
due to spin ($\Delta \langle s_x \rangle_i$) and orbital ($\Delta \langle l_x \rangle_i$) 
angular momenta using equations similar to Eq.\ \ref{eq.ch8:21}. As shown in right panels
of Fig.\ \ref{fig.ch8:2} these contributions have complicated dependences both on the 
frequency and the structure of the single-particle orbital under study \cite{AR.00}.

 Oscillator shells with higher $N$-values show a similar behavior. At
a given rotational frequency the modifications of single-particle
$\langle \hat{\jmath}_{x} \rangle _{i}$ values induced by NM lead to 
an increase of the total $J$ and of the moments of inertia (see Figs.\
\ref{fig.ch8:3} and \ref{fig.ch8:5} below). This means 
that NM enhances the angular momentum in rotating nuclei in addition 
to the Coriolis term.

\begin{figure}
\centering
\includegraphics[width=10.0cm]{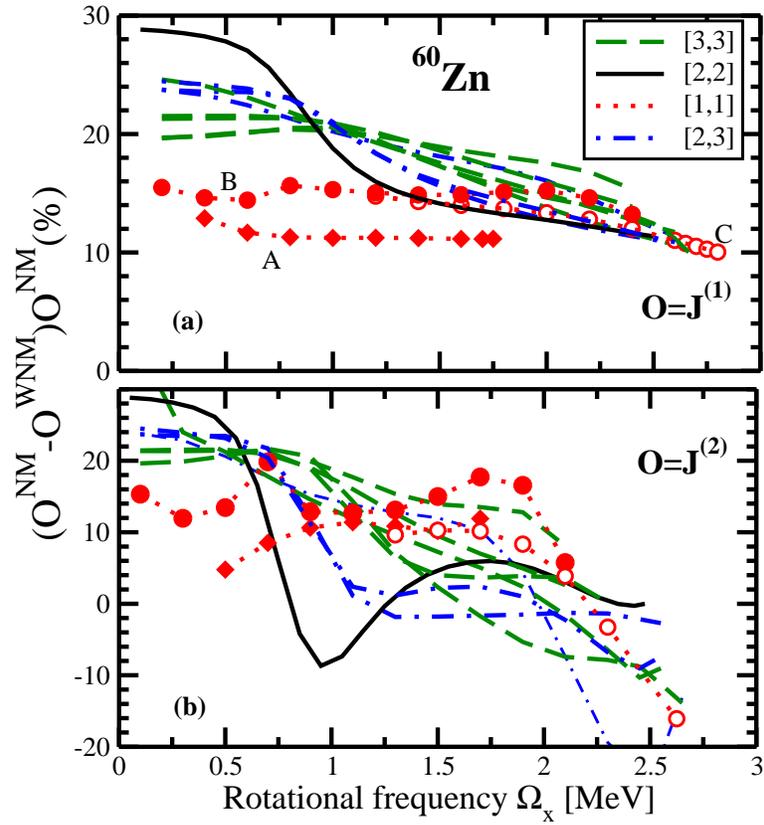}
\caption{The contributions of NM to the dynamic ($J^{(2)}$) (panel (b)) 
and kinematic ($J^{(1)}$) (panel (a)) moments of inertia as a function of 
rotational frequency for highly-deformed and superdeformed configurations 
of $^{60}$Zn obtained in the CRMF calculations. Different color/line types 
are used for different groups of configurations.
notation $[n,p]$, where $n(p)$ is the number of occupied $g_{9/2}$
neutrons (protons). From Ref.\ \cite{Afanasjev2010Phys.Rev.C34329}.
\label{fig.ch8:3}}
\end{figure}

\begin{figure}
\centering
\includegraphics[width=12.0cm]{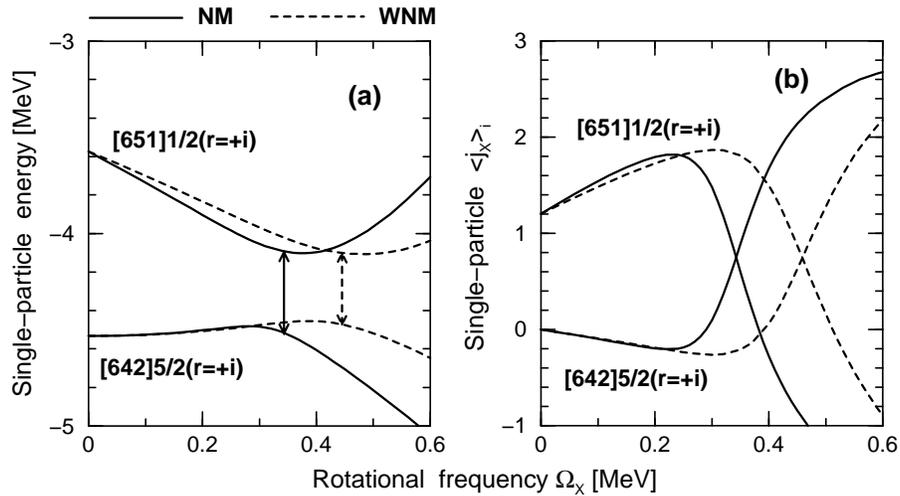}
\caption{(a) Proton single-particle energies (Routhians)
in the self-consistent rotating potential as a function of
rotational frequency $\Omega_x$ obtained in the CRMF
calculations with and without NM. They are given along the
deformation path of the lowest SD configuration in $^{194}$Pb.
Only interacting $[651]1/2^+$ and $[642]5/2^+$ orbitals
are shown, see Fig.\ 1 in Ref.\ \protect\cite{Afanasjev2000Nucl.Phys.A196} for full
spectra. (b) The expectation values $\langle \hat{j}_x \rangle_i$
of the single-particle angular momentum operator $\hat{j}_x$ of the
orbitals shown on panel (a). Solid and dashed arrows are used to 
indicate the frequencies  (as well as the energy gap between the 
interacting orbitals in panel (a)) at which the band crossings take 
place  in the calculations with and without NM, respectively.
From Ref.\ \cite{Afanasjev2010Phys.Rev.C34329}.
\label{fig.ch8:4}}
\end{figure}
 
\begin{figure}
\centering
\includegraphics[width=15.0cm]{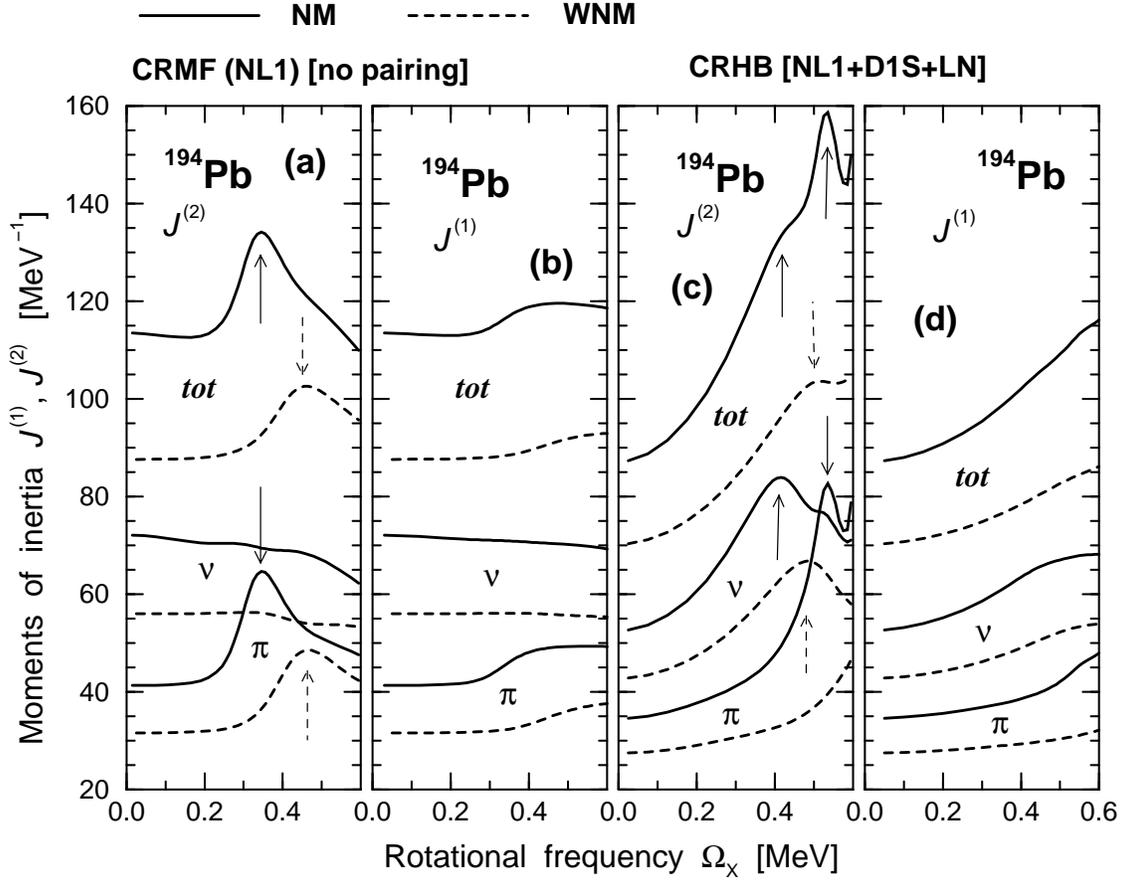}
\caption{Kinematic ($J^{(1)}$) and dynamic ($J^{(2)}$) moments of inertia for the 
lowest SD configuration in $^{194}$Pb obtained in the calculations with and without 
NM. Proton and neutron contributions to these quantities are indicated by $\pi$ and 
$\nu$, while total moments by '$tot$'. Panels (a) and (b) show the results obtained in 
the calculations without pairing, while panels (c) and (d) show the results of the 
calculations within the CRHB+LN framework. Solid and dashed arrows are used to indicate 
the frequencies at which the band crossings take place in the calculations with and 
without NM, respectively. From Ref.\ \cite{Afanasjev2010Phys.Rev.C34329}.
\label{fig.ch8:5}}
\end{figure}

  The most important impact of NM on physical observables is revealed in the 
moments of inertia. To quantify this impact, the contribution $\Delta O^{NM-contr}$ 
(in percentage) of NM to the physical observable $O$ is defined as
\begin{eqnarray}
\Delta O^{NM-contr} = \frac{O^{NM}-O^{WNM}}{O^{NM}}\times 100\%.
\label{eq.ch8:22}
\end{eqnarray} 

  The increase of the moments of inertia due to NM has been studied 
first in a semiclassical approximation in Ref.\ \cite{Koepf1990Nucl.Phys.A279}, then in 
fully self-consistent CRMF calculations on the example of the yrast SD band 
in $^{152}$Dy in Ref.\ \cite{Konig1993Phys.Rev.Lett.3079}  and finally in a systematic way in 
Refs.\ \cite{AR.00,A.08,Afanasjev2010Phys.Rev.C34329}. NM  typically increases the calculated 
kinematic moments of inertia of normal-deformed rotational bands in 
the rare-earth region by  10-30\% \cite{Afanasjev2010Phys.Rev.C34329}. Considerable fluctuations 
of the $(J^{(1)}_{NM}-J^{(1)}_{WNM})/J^{(1)}_{NM}$ quantity as a function of proton 
and neutron numbers seen in some isotonic and isotopic chains at normal 
deformation are due to the changes in underlying single-particle 
structure. The $(J^{(1)}_{NM}-J^{(1)}_{WNM})/J^{(1)}_{NM}$ quantity is 
around 20\% in the superdeformed bands of the $A\sim 150$ mass
region and around 20-25\% in the hyperdeformed bands in the 
$Z=40-58$ part of nuclear chart  at typical frequencies at which
these bands are either observed or expected to be observed.

  The configuration and frequency dependence of the impact of NM on 
the moments of inertia is shown in Fig.\ \ref{fig.ch8:3}a.
With increasing rotational frequency, the average contribution of NM 
into kinematic moments of inertia decreases and it falls below 15\% at 
$\Omega_x\sim 2.5$ MeV . In addition, the configuration dependence of the 
$\Delta J^{(1)}_{NM-contr}$ quantities is weaker than the one at low 
frequencies. At these frequencies, the majority of occupied single-particle 
orbitals are either completely aligned or very close to complete alignment. 
However, NM do not modify the expectation values  of the single-particle 
angular momenta $\left< j_x \right>_i$  of completely aligned orbitals 
\cite{A.08}. As a result, only remaining orbitals, which are still aligning, 
contribute into $\Delta J^{(1)}_{NM-contr}$. The combined contribution of these 
orbitals into $\Delta J^{(1)}_{NM-contr}$ is smaller than the one at lower 
frequencies because the alignment of these orbitals is not far away from 
complete. The impact of NM on the dynamic moments of inertia is shown in 
Fig.\ \ref{fig.ch8:3}b and it clearly displays much more complicated 
pattern as compared with the impact of NM on the kinematic moments of inertia
(see Ref.\ \cite{Afanasjev2010Phys.Rev.C34329} for details).

  The modification of the single-particle alignments and energies in the 
presence of NM leads to substantial impact on the band crossing properties. 
This is illustrated in Figs.\ \ref{fig.ch8:4} and \ref{fig.ch8:5} on the 
example  of the lowest superdeformed (SD) band in $^{194}$Pb.
  
  In the CRMF calculations, the unpaired proton band crossing originates 
from the interaction between the $\pi[642]5/2^+$ and $\pi[651]1/2^+$ orbitals 
(Fig.\ \ref{fig.ch8:4}a). Since NM increases somewhat the single-particle 
alignment $\langle \hat{j}_x \rangle_i$ (Fig.\ \ref{fig.ch8:4}b) and the 
slope of the routhian for the $\pi[651]1/2^+$ orbital (Fig.\ \ref{fig.ch8:4}a), 
the band crossing takes place at lower frequency. The shift of crossing frequency 
due to NM is considerable (120 keV) from 0.465 MeV (WNM) down to 0.345 MeV (NM), 
Fig.\ \ref{fig.ch8:4}a. The calculations also suggest that the strength of the 
interaction between two interacting orbitals at the band crossing is modified in 
the presence of NM as seen in the change of the energy distance (gap) between 
these two orbitals at the crossing frequency (Fig.\ \ref{fig.ch8:4}a).

  The impact of NM on band crossing features is also seen in the CRHB+LN calculations 
where the alignment of the pairs of $j_{15/2}$ neutrons and $i_{13/2}$ protons causes 
the shoulder and peak in total dynamic moment of inertia $J^{(2)}$ (Fig.\ 
\ref{fig.ch8:5}c) (see also Ref.\ \cite{Afanasjev2000Nucl.Phys.A196}). Note that each of these two 
alignments creates a peak in the dynamic moment of inertia of corresponding subsystem. 
NM shifts the paired neutron band crossing to lower frequencies by 70 keV from $0.485$ 
MeV (WNM) to $0.415$ MeV (NM). Paired proton band crossing lies in the calculations 
with NM at $\Omega_x=0.535$ MeV, while only the beginning of this crossing is seen 
in the calculations without NM (Fig.\ \ref{fig.ch8:5}c).

  The origin of this effect is twofold. Similar to the  unpaired calculations, 
the part of it can be traced to the fact that NM increases the expectation values
$\langle \hat{j}_x \rangle_i$ of the orbitals located at the bottom of the shell 
(the discussed orbitals are of this kind) \cite{AR.00}. The corresponding larger 
slope of the quasiparticle routhians causes the shift of the crossing to lower 
frequencies. 

  However, an additional contribution comes from the modification of the 
pairing  by NM. There is a difference in the pairing energies calculated with 
and without NM which increases with rotational frequency, see Fig.\ 3c in Ref.\ 
\cite{Afanasjev2010Phys.Rev.C34329}. The pairing in the calculations with NM is weaker. This can be  
explained by the increase of $\langle \hat{j}_x \rangle_i$ of the orbitals 
located at the bottom of the shell due to NM (see above). Thus in the presence 
of NM the gradual breaking of high-$j$ pairs proceeds faster which leads to a 
faster reduction of pairing with increasing $\Omega_x$ as compared with the WNM 
calculations. This effect is called as {\it an anti-pairing  effect induced by 
NM} \cite{Afanasjev2010Phys.Rev.C34329}. 

   Above discussed CRMF and CRHB+LN examples clearly show that the modifications 
of band crossing  features (crossing frequencies and the features of the kinematic 
and dynamic moments of inertia in band crossing region) caused by NM are substantial 
and depend on the underlying modifications of single-particle properties such 
as alignments and single-particle (quasi-particle) energies.

 The changes of single-particle properties induced by nuclear magnetism affect 
also other physical observables in rotating nuclei, the discussion of which is 
not possible due to space limitations. These are
\begin{itemize}
\item {\it modifications of the energy splittings between signature partner
orbitals (signature splitting)} \cite{AR.00}. These modifications 
represent an additional source of band crossing frequency changes in odd-
and odd-odd nuclei and excited configurations of even-even ones in the
presence of NM (Sect. III in Ref.\ \cite{Afanasjev2010Phys.Rev.C34329}.

\item {\it modifications of the effective alignments $i_{eff}$ \cite{AR.00}.} 
This observable is often used in the analysis of the single-particle structure 
of superdeformed bands (see Ref.\ \cite{ALR.98})

\item
{\it the existence of signature-separated rotational bands.} 
\cite{MDD.00,Pingst-A30-60,Afanasjev2010Phys.Rev.C34329}.
They reveal themselves in a considerable energy splitting of the $r_{tot}=+1$ and $r_{tot}=-1$ 
branches of the configurations which have the same structure in terms of occupation 
of single-particle states with given Nilsson labels.  {\it This feature is a 
strong spectroscopic fingerprint of the presence of time-odd mean fields.}

\item
{\it Within specific configuration the impact of NM on the binding energies
reaches its maximum at the terminating state \cite{A.08}.} Underlying 
microscopic mechanism for additional binding due to NM at such states 
has the same features as those seen in low-spin one- and two-particle 
configurations of odd and odd-odd nuclei \cite{Afanasjev2010Phys.Rev.C14309}. However, the 
magnitude of the effects is significantly larger. 

\item
 The values of kinematic moment of inertia calculated with 
NM are typically within 5\% of the rigid body value for the moments of 
inertia at super- and hyperdeformation \cite{Afanasjev2010Phys.Rev.C34329}, but the deviations 
from the rigid-body value are significantly larger for normal-deformed bands. 

\item
   NM has very small effect on the deformation properties of nuclei 
\cite{AR.00,Afanasjev2010Phys.Rev.C14309,Afanasjev2010Phys.Rev.C34329}.

\end{itemize}

\section{Currents in the intrinsic (rotating) frame}

\begin{figure}[th]
\centering
\includegraphics[width=15.5cm]{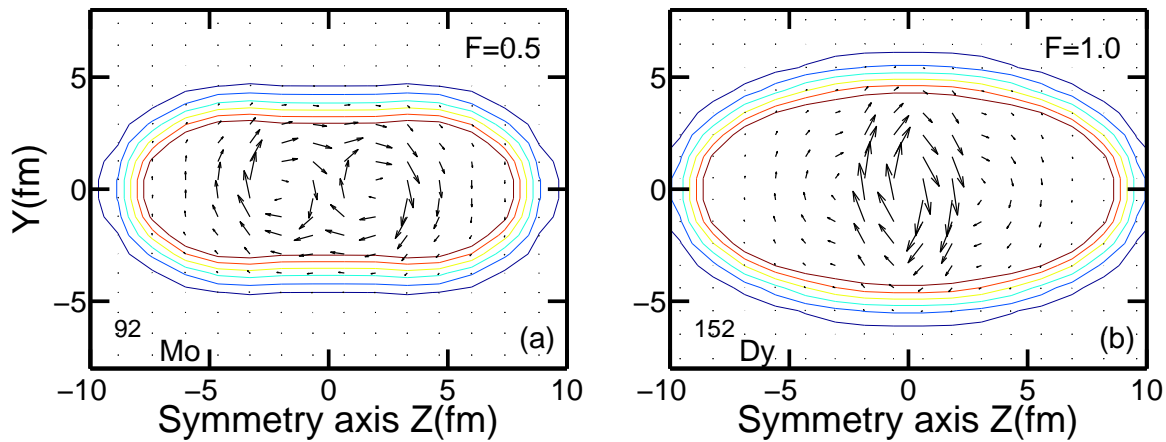}
\caption{Total neutron current distributions  {\bf j}$^n$({\bf r}) in the 
intrinsic frame in the $y-z$ plane for the yrast hyperdeformed ($^{92}$Mo) 
and superdeformed ($^{152}$Dy) configurations at rotational frequencies 
$\Omega_x \sim 1.0$ MeV and $\Omega_x=0.5$ MeV, respectively. The currents 
in panel (b) are plotted  at arbitrary units for better visualization. The 
currents in panel (a) are normalized to the currents in panel (b) by using 
factor F. The shape and size of the nucleus are indicated by density lines 
which are plotted in the range $0.01-0.06$ fm$^{-3}$ in step of 0.01 fm$^{-3}$. 
Based on Fig.\ 8 of Ref.\ \cite{Afanasjev2010Phys.Rev.C34329}.
\label{fig.ch8:6}}
\end{figure}

  The Coriolis term is present in NM and WNM calculations. This means 
that the currents (Eq.\ (\ref{eq.ch8:10})) are always present in rotating 
nuclei. However, it is important to distinguish the currents induced by the 
Coriolis term and the ones which appear due to magnetic potential. The currents, 
which appear in the WNM calculations, are generated by the Coriolis term. Thus, 
following Ref.\ \cite{Afanasjev2010Phys.Rev.C34329} they are called as {\it Coriolis induced currents}.  
On the contrary, the currents in the NM calculations are generated by both the 
Coriolis term and magnetic potential. The difference of the currents in the NM 
and WNM calculations is attributable to magnetic potential. Thus,  the currents 
$[{\bf j}^{n,p}({\bf r})]^{NM} - [{\bf j}^{n,p}({\bf r})]^{WNM}$ are called as 
{\it magnetic potential induced currents} \cite{Afanasjev2010Phys.Rev.C34329}.

\begin{figure}
\centering
\includegraphics[width=17.5cm]{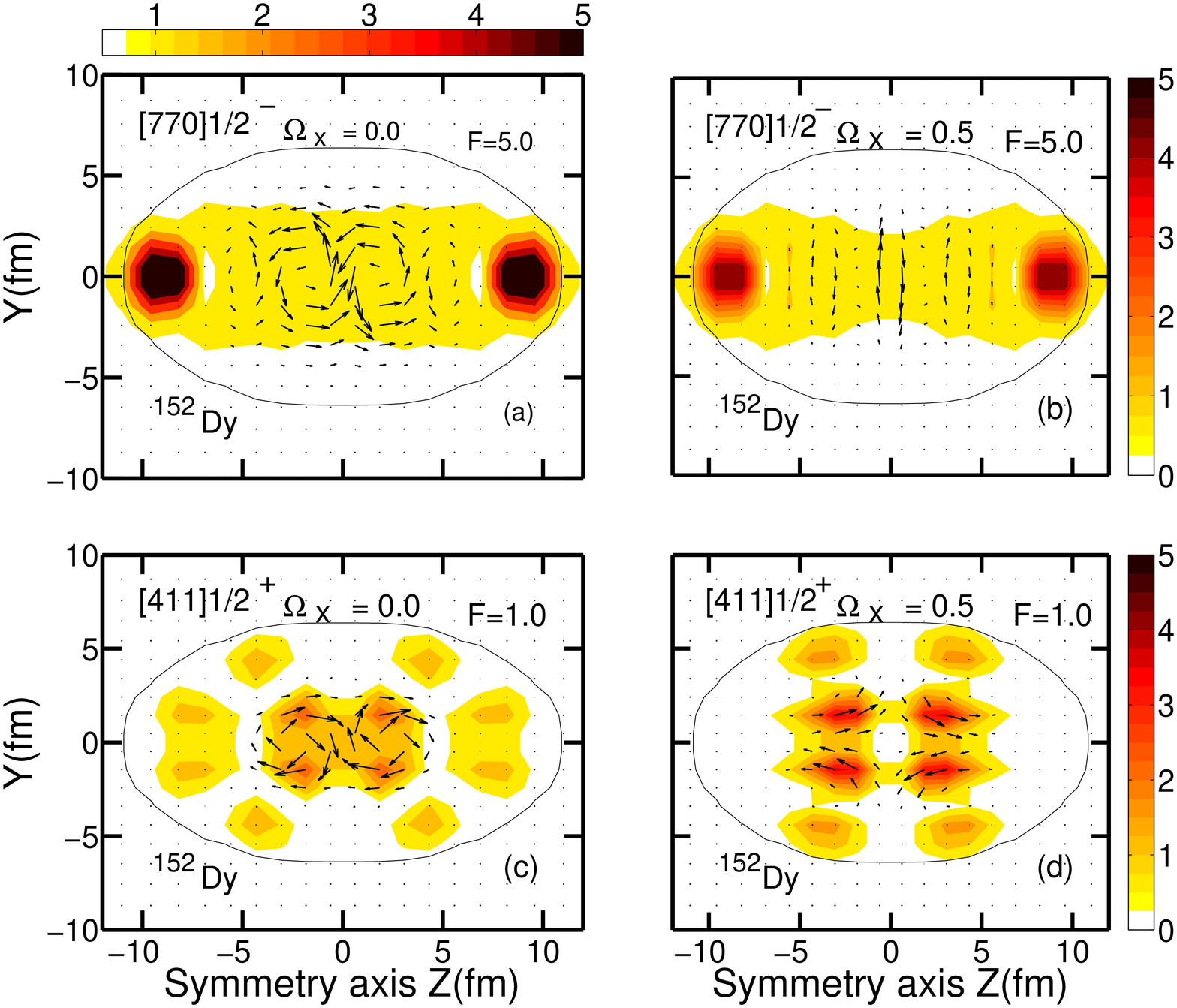}
\caption{Current distributions {\bf j}$^n$({\bf r}) produced by 
single neutron in indicated single-particle states of the yrast SD configuration in 
$^{152}$Dy at rotational frequencies $\Omega_x=0.0$ MeV (left panels) and $\Omega_x=0.5$ 
MeV (right panels). The shape and size of the nucleus are indicated by density line 
which is plotted at $\rho=0.01$ fm$^{-3}$. The currents in panels (c) and (d) are plotted at 
arbitrary units for better visualization. The currents in panels (a) and (b) are normalized to 
the currents in panels (c) and (d) by using factor F.  The currents and densities 
are shown in the intrinsic frame in the $y-z$ plane at $x=0.48$ fm. The single-neutron density 
distributions due to the occupation of the indicated Nilsson state are shown by colormap. 
Note that slightly different colormap is used in panel (a) for better visualization of 
densities. Based on Fig.\ 9 of Ref.\ \cite{Afanasjev2010Phys.Rev.C34329}.
\label{fig.ch8:7}}
\end{figure}
 
   The detailed analysis of the currents in the rotating nuclei in the CDFT
framework has first been performed in Ref.\ \cite{Afanasjev2010Phys.Rev.C34329} \footnote{For 
earlier studies of the currents in rotating nuclei in non-relativistic frameworks 
see Refs.\ \cite{KG.77,DSK.85,MQS.97,LSQM.03,R.76,GR.78-0,GR.78,KM.79,FKMW.80}.} 
The currents in the rotating frame of reference that is fixed to the body is caused 
by quantized motion of the fermions. Complicated structure  of the currents in the 
rotating systems of independent fermions visible in Fig.\ \ref{fig.ch8:6} is the 
consequence of the fact that total current is the sum of the single-particle currents. 
The single-particle currents show vortices  (circulations), the  existence of which 
implies non-vanishing current circulations defined as 
$\bf{C}(\bf{r})=\bf{\nabla} \times \bf{j}(\bf{r})$ 
\cite{GR.78}. Note that the strength and localization of vortices depends on the
single-particle state (see Fig.\ \ref{fig.ch8:7}). As a result, the differences
seen in total currents of different nuclei/structures in Fig.\ \ref{fig.ch8:6} can 
be traced back to the single-particle content of underlying single-particle
configuration.

  The localization, the strength and the structure of the current vortices created 
by a particle in a specific single-particle state depend on its nodal structure (see 
Ref.\ \cite{GR.78} and Sec.\ IIIC in Ref.\ \cite{Afanasjev2010Phys.Rev.C14309}). In a slowly rotating 
anisotropic harmonic oscillator potential Coriolis induced current for a single particle 
shows a rather simple structure with the centers of the circulations found at the nodes 
and peaks of the oscillator eigenfunctions \cite{GR.78}; this structure forms a rectangular 
array somewhat similar to a crystal lattice. The analysis of single-particle vortices in 
rotating nuclei in the CRMF framework  in general confirms these results (see Fig.\ 
\ref{fig.ch8:7} and Ref.\ \cite{Afanasjev2010Phys.Rev.C34329}). In part, this is due to the fact that 
magnetic potential induced currents are weaker than Coriolis induced ones.

  The typical features of the single-particle currents in the CRMF approach are 
seen in Fig.\ \ref{fig.ch8:7}. The comparison of left and right columns of this 
figure clearly indicates  that for a given single-particle state the increase of 
rotational frequency (i) does not lead to appreciable modifications of the density 
distribution  but (ii) considerably modifies the strength of the currents and
changes the shape of circulations. The latter is due to two factors. First,
the Coriolis induced currents become active at $\Omega_x \neq 0.0$ MeV and 
at $\Omega_x=0.5$ they are dominant type of currents. Second, the wave function 
undergoes considerable modifications with increasing rotational frequency.
For example, the wave function (in terms of two largest components) of the 
$\nu [770]1/2^-$ state changes from 62\%[770]1/2+17\%[761]1/2  at $\Omega_x=0.0$ 
MeV to 39\%[770]1/2+28\%[761]3/2  and $\Omega_x=0.5$ MeV.

  The total current is the sum of Coriolis induced and magnetic potential induced 
currents. In the majority of the cases total current is dominated by the Coriolis 
induced currents; magnetic potential induced currents represent approximately 5-20\% 
[30\%] of total current in the HD and SD  [ND] nuclei \cite{Afanasjev2010Phys.Rev.C34329}. The spatial 
distribution of Coriolis induced and magnetic potential induced currents is similar 
in the majority of nuclei. However, there are cases in which the spatial distribution 
of these two types of currents differ substantially \cite{Afanasjev2010Phys.Rev.C34329}. Note that current
is weak in the surface area (Fig.\ \ref{fig.ch8:6}). This is contrary to semiclassical 
description of currents in normal and superfluid rotating nuclei \cite{DSK.85} 
according to which the average intrinsic current flows mainly in the nuclear surface
area. This underlines the importance of quantum mechanical treatment of the currents.
 
  It is well known that there are no currents in the intrinsic frame if the rigid 
non-spherical body rotates uniformly (rigid rotation) (see Sec.\ 6A-5 in Ref.\ 
\cite{Bohr1975}).  The moments of inertia of super- and hyperdeformed configurations 
in unpaired regime come very close to the rigid-body values
in the CRMF calculations \cite{Afanasjev2010Phys.Rev.C34329}. However, the intrinsic currents display
the dramatic deviations from rigid rotation (Fig.\ \ref{fig.ch8:6}).  This clearly 
shows that the closeness of the moments of inertia to rigid body value does not 
necessary implies that the current distribution should correspond to rigid rotation. 
On a microscopic level, the building blocks of the total current, namely, the 
single-particle currents certainly do not have a rigid-flow character; on the contrary, 
they have the vortex-flow character (see Fig.\ \ref{fig.ch8:7}).

\section{Ground state rotational bands in normal-deformed even-even nuclei}

 Despite considerable amount of experimental data on normal-deformed
nuclei and the availability of relevant theoretical frameworks, only 
recently and only in a single region (actinides) of nuclear chart the 
systematic investigation of rotational properties has been performed 
in the DFT framework. These are the studies of Refs.\ \cite{AO.13,A.14} 
performed in the CDFT framework. On the contrary, only few nuclei have
been studied at normal deformation in non-relativistic DFT 
\cite{ER.94,AER.01,AER.01b,DBH.01,BBDH.03}. In the current section I
will review the major features of the rotational bands in the actinides
and assess the accuracy of the description of these bands in the CRHB+LN 
framework.

\begin{figure*}[ht]
\centering
\includegraphics[width=17.5cm,angle=0]{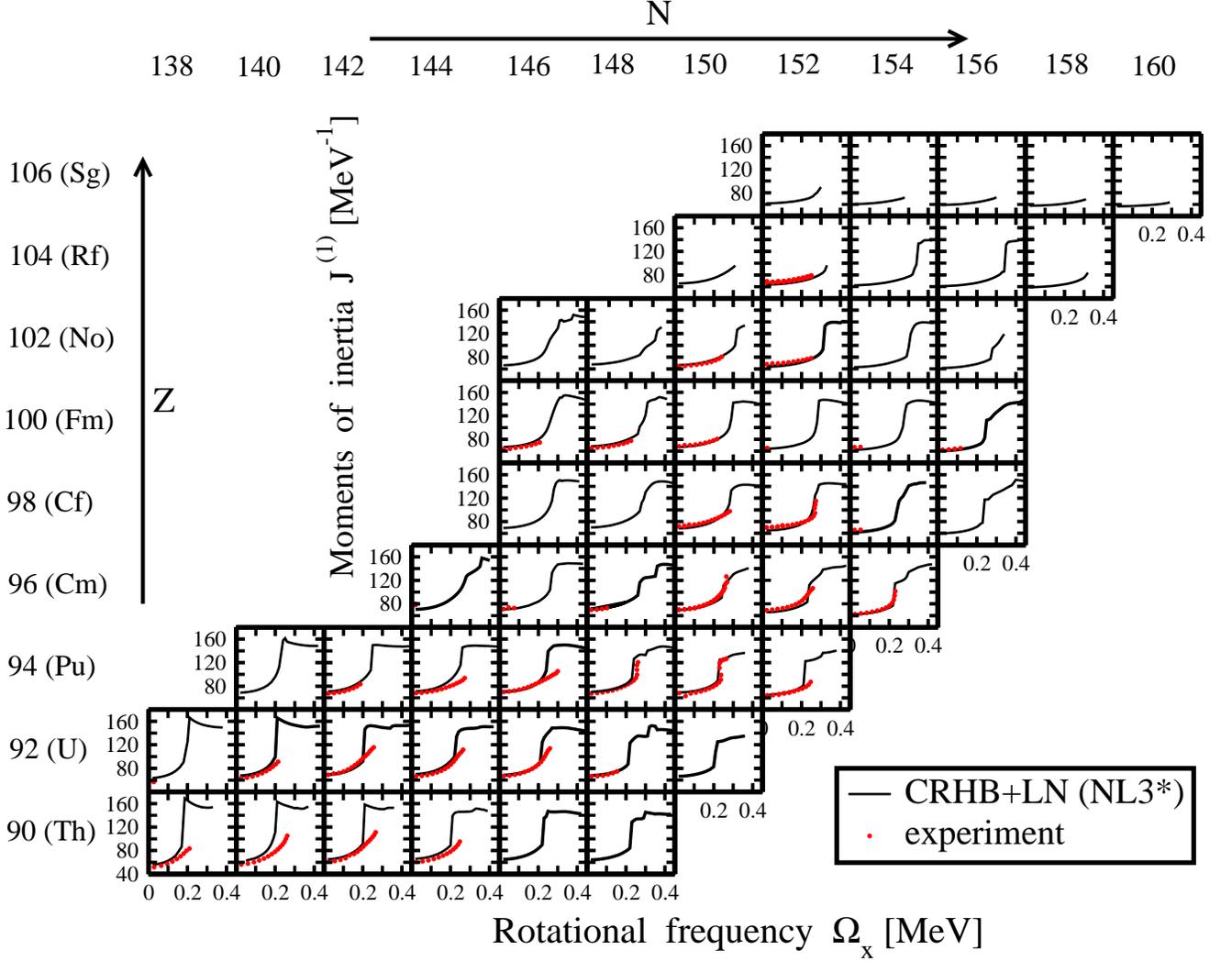}
\caption{The experimental and calculated moments of inertia $J^{(1)}$ 
as a function of rotational frequency $\Omega_x$. The calculations 
are performed with the NL3* CEDF \cite{Lalazissis2009Phys.Lett.B36}. Calculated results 
and experimental data are shown by black lines and red dots, 
respectively. Cyan dots show new experimental data from Ref.\ \cite{Hota.thesis} 
which were not included in Ref.\ \cite{AO.13}.
From Refs.\ \cite{AO.13,A.14}.} 
\label{fig.ch8:8}
\end{figure*}

  Fig.\ \ref{fig.ch8:8} shows the results of the first ever (in 
any DFT framework) systematic investigation of rotational properties 
of even-even nuclei at normal deformation \cite{AO.13}. The calculations
are performed within the CRHB+LN approach \cite{Afanasjev2000Nucl.Phys.A196}. One can see that 
the gradual increases of the moments of inertia below band crossings 
are reproduced well. The upbendings observed in a number of rotational 
bands of the $A\geq 242$ nuclei are also reasonably well described in model 
calculations. However, the calculations also predict similar upbendings 
in lighter nuclei, but they have not been seen in experiment. Note that similar
problems exist also in the cranking model calculations based on 
phenomenological potentials \cite{AO.13}. The analysis of Ref.\ 
\cite{AO.13} suggests that the stabilization  of octupole deformation 
at high spin, not included in the present CRHB+LN calculations, could 
be responsible for this discrepancy between theory and experiment.

\begin{figure}
\centering
\includegraphics[width=10.5cm,angle=0]{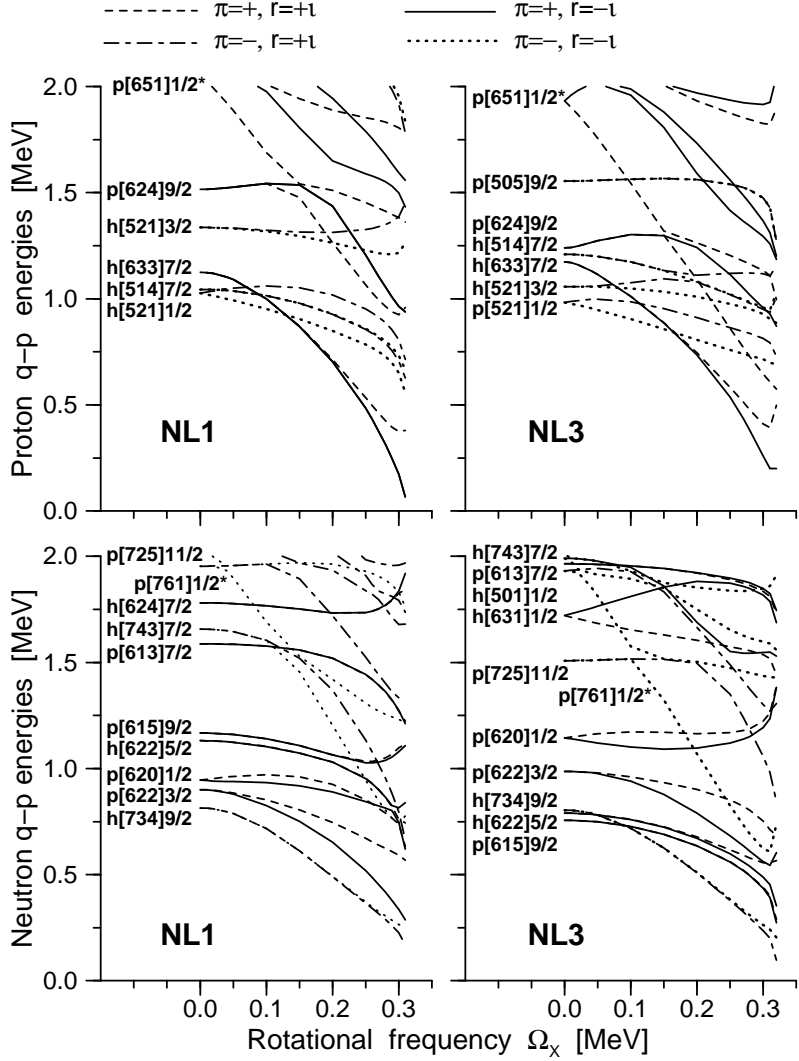}
\caption{Proton (top panels) and neutron (bottom panels)
quasiparticle energies corresponding to the lowest configuration
in $^{254}$No. The CRHB+LN calculations have been performed 
with the NL1 (left panels) and NL3 (right panels) parametrizations. 
The letters 'p' and 'h' before the Nilsson labels are used to 
indicate whether a given routhian is of particle or hole type.
From Ref.\ \cite{A250}.} 
\label{fig.ch8:9}
\end{figure}

\begin{figure}
\centering
\includegraphics[width=12.5cm,angle=0]{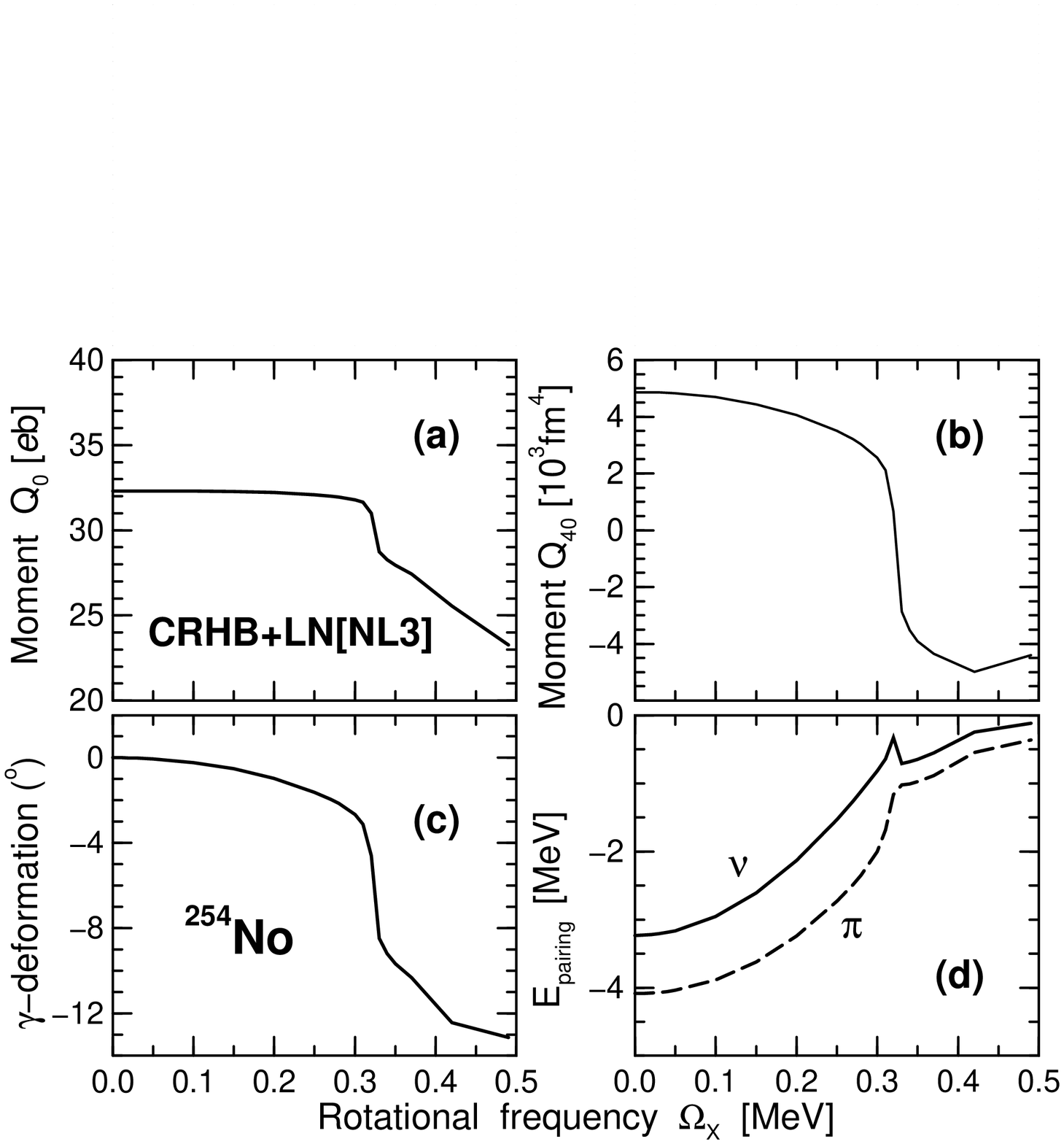}  
\caption{Calculated mass quadrupole ($Q_0$) and hexadecapole 
($Q_{40}$) moments, the $\gamma$-deformation and proton and 
neutron pairing energies ($E_{pairing}$) of the ground state
rotational band in $^{254}$No as a function of rotational 
frequency $\Omega_x$. The results are obtained in the CRHB+LN 
calculations with the NL3 CEDF. From Ref.\ \cite{A250}.}
\label{fig.ch8:10}
\end{figure}

 Either sharp or more gradual increases of the kinematic moments of inertia 
are observed at $\Omega_x \approx 0.2-0.30$ MeV (Fig.\ \ref{fig.ch8:8}). They 
are due to the alignments of the neutron $j_{15/2}$ and proton $i_{13/2}$ orbitals 
which in many cases take place at similar rotational frequencies. The situation 
is more complicated than in the rare-earth region in which the $h_{11/2}$ protons 
align substantially later than the $i_{13/2}$ neutrons. For example, 
the routhian diagrams (Fig.\ \ref{fig.ch8:9}) show simultaneous alignment 
of the proton $i_{13/2}$ pair ($\pi[633]7/2$ Nilsson orbit) and neutron $j_{15/2}$ 
pair ($\nu[734]9/2$ orbit) at $\Omega_x\approx 0.32$ MeV in $^{254}$No. In the 
calculations with NL3 CEDF, the total angular momentum gain at the band crossing 
is $\approx 17\hbar$, with proton and neutron contributions of 
$\approx 7\hbar$ and $\approx 10\hbar$, respectively \cite{A250}. The alignment 
of these orbitals leads to a decrease of the mass hexadecapole moment $Q_0$, to 
a sign change of the mass hexadecapole moment $Q_{40}$ and to an appreciable 
increase of the absolute value of the $\gamma$-deformation (Fig.\ \ref{fig.ch8:10}a,b and c); 
this type of behavior is typical for the bands in this mass region. In addition, the 
pairing energies decrease with increasing rotational frequency due to the Coriolis 
anti-pairing effect and the pairing becomes very weak above the band crossing region 
(Fig.\ \ref{fig.ch8:10}d).

\begin{figure}[ht]
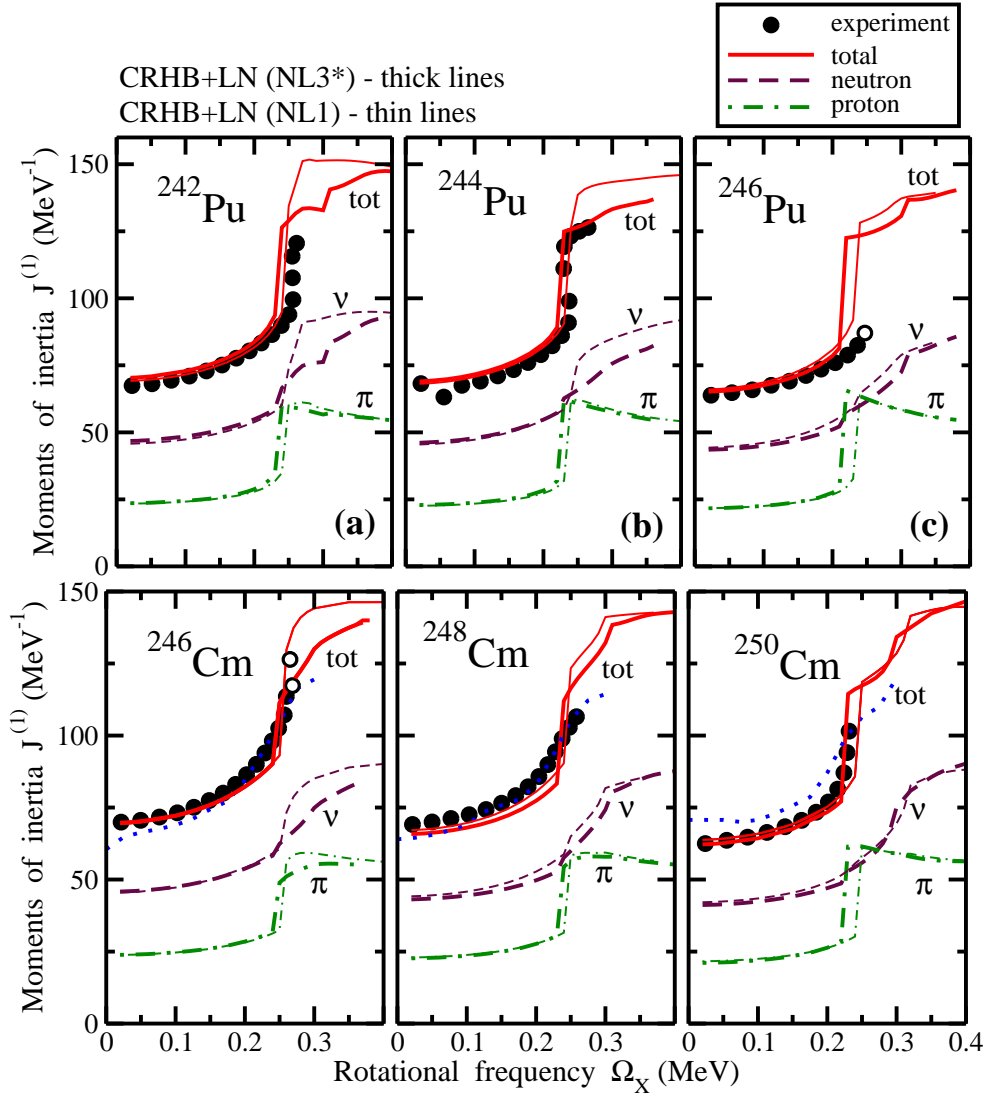

\centering
\includegraphics[width=12.7cm]{fig-11-top.eps}
\includegraphics[width=13.0cm]{fig-11-bot.eps}
\caption{
The experimental and calculated kinematic moments of inertia $J^{(1)}$ 
of ground state rotational bands in indicated nuclei as a function of 
rotational frequency $\Omega_x$. Proton and neutron contributions to the 
kinematic moment of inerta are presented. Open circles are used for 
tentative experimental points. Total kinematic moments of inertia 
obtained in the CSM+PNC approach \cite{ZHZZZ.12} are shown by blue solid 
lines. Based on Ref.\  \cite{A.14}.
} 
\label{fig.ch8:11}
\end{figure}

  Fig.\ \ref{fig.ch8:11} illustrates the accuracy of the description of band 
crossing  features in the Pu and Cm nuclei. Here I will discuss in detail only 
the band crossings in the Pu isotopes. One can see that apart of $^{250}$Cm rather 
good description of experimental data has been obtained in the CRHB+LN 
calculations.

  The CRHB+LN calculations predict sharp upbend $\Omega_x\sim 0.25$ MeV
in all three even-even Pu isotopes \cite{AO.13}. The upbending is complete 
in $^{244}$Pu and the CRHB+LN(NL3*) 
calculations rather well describe it  (Fig.\ref{fig.ch8:11}b); the sharp 
alignment of the proton $i_{13/2}$ orbitals is a source of this backbending 
and the neutron $j_{15/2}$ alignment proceeds gradually over extended frequency 
range. On the contrary, sharp alignments of the proton and neutron pairs take 
place at the same frequency in the CRHB+LN(NL1) calculations  (Fig.\ 
\ref{fig.ch8:11}b) and they somewhat overestimate the kinematic moment of inertia 
above the band crossing. The same situation with the alignments of the proton 
$i_{13/2}$ and neutron $j_{15/2}$ pairs exists also in the CRHB+LN(NL1) and 
CRHB+LN(NL3*) calculations for  $^{242}$Pu. They accurately reproduce the 
evolution of kinematic moments of inertia with frequency and the frequency of 
the paired band crossing (Figs.\ \ref{fig.ch8:8} and \ref{fig.ch8:11}a). However, 
since upbending is not complete in experiment it is impossible 
to judge whether the simultaneous sharp alignments of proton 
and neutron pairs really take place  in nature. The kinematic moment of inertia 
of the ground state rotational band in  $^{246}$Pu shows a rapid increase at the 
highest observed frequencies similar to the one seen before upbendings in 
$^{242,244}$Pu. However, the $^{246}$Pu data does not reveal an upbend yet. 
The upbend in $^{242,244}$Pu is predicted $0.01-0.02$ MeV earlier in the 
CRHB+LN(NL3*) calculations as compared with experiment. A similar situation
is expected in $^{246}$Pu. Considering this and the fact that the last 
observed point in $^{246}$Pu is tentative, one can conclude that 
there is no significant discrepancies with experimental data. Even 
better agreement with this new data is seen in the case of the 
CRHB+LN(NL1) calculations.

  Similar CRHB+LN analysis of the paired band crossings in $^{248,250}$Cf and
$^{246,248,250}$Cf (see also Figs.\ \ref{fig.ch8:11}d,e and f) has been presented
in Ref.\ \cite{A.14}. In this reference, the CRHB+LN results for the Cf and 
Cm nuclei were compared with the ones obtained in the cranked shell model with 
the pairing correlations treated by a particle-number conserving method (further 
CSM+PNC) \cite{ZHZZZ.12}. There are three important differences between the 
CRHB+LN and CSM+PNC approaches. First, the parameters of the Nilsson potential 
were carefully adjusted in the CSM+PNC approach to the experimental energies 
of deformed one-quasiparticle states of actinides in Ref.\ \cite{ZHZZZ.12}.
Second, in the CSM+PNC approach the deformations are chosen to be close to 
experimental values and they do not change with rotational frequency.
These two types of observables are defined fully self-consistently in
the CRHB+LN approach without any fit to experimental data, and, in addition,
the deformations (mass moments) change with rotational frequency (Fig.\ 
\ref{fig.ch8:10}d). Third, the pairing strength has been fitted to 
experiment in both approaches \cite{AO.13,ZHZZZ.12}. However, the pairing 
strength is different in even-even and odd-mass nuclei in the CSM+PNC approach 
\cite{ZHZZZ.12}; this is a well known deficiency of the cranked shell model (see, 
for example, Ref.\ \cite{CSM-exp}). In contrast, the same pairing strength is 
used in even-even and odd-mass nuclei in the CRHB+LN approach; it leads to a 
consistent and accurate description of odd-even mass staggerings (the $\Delta^{(3)}$ 
indicators) and the moments of inertia in even-even and odd-mass actinides
(see Ref.\ \cite{AO.13} and Sect.\ \ref{rot-odd}).

  Thus, CRHB+LN approach provides much more consistent description of 
rotational properties in paired regime as compared with the CSM+PNC 
approach; with no adjustment of single-particle energies and deformations 
to experiment it obtains comparable in accuracy description of experimental 
rotational properties of actinides \cite{A.14}. In addition to these 
studies in actinides, the CRHB+LN approach has succefully been applied to 
the investigations of the ground state rotational bands in the $A\sim 70$ $N\sim Z$ 
region \cite{AF.05} and of the evolutions of the moments of inertia  
with particle numbers at low frequency ($\Omega_x=0.02$ MeV) in the 
rare-earth region \cite{J1Rare}.

\section{Rotational bands of odd-mass nuclei in paired regime}
\label{rot-odd}

\begin{figure}[ht]
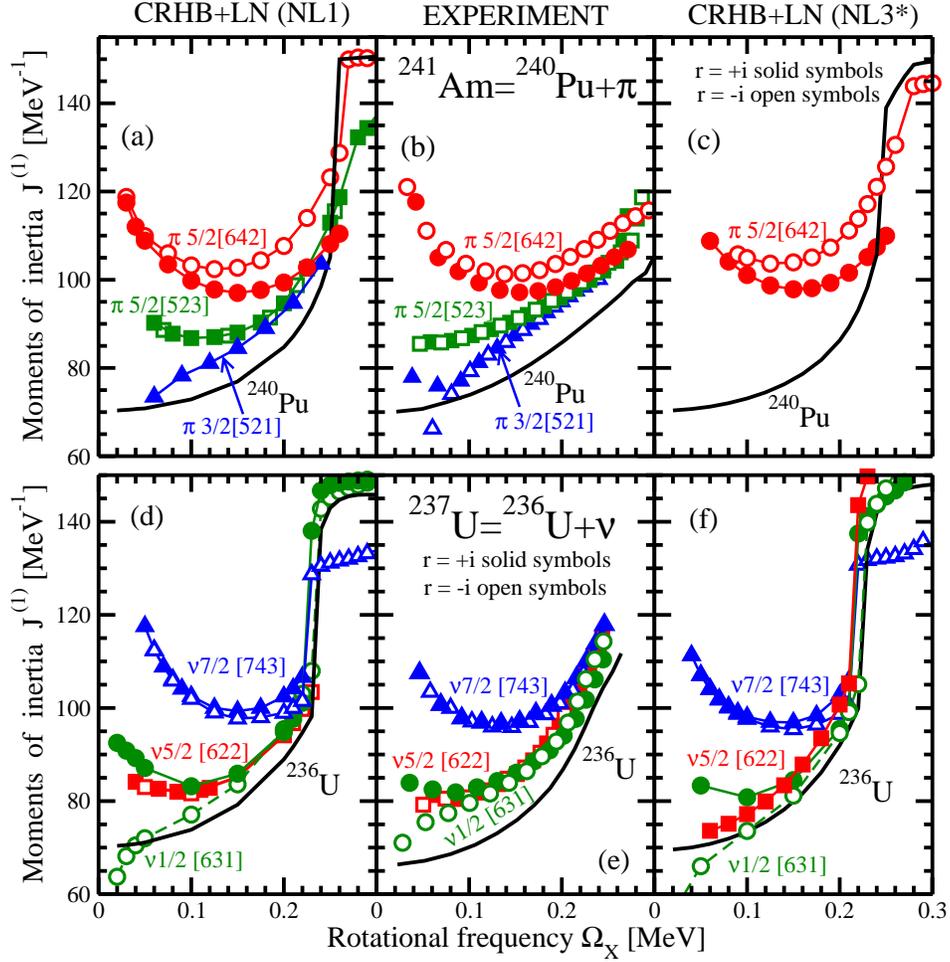

\centering
\includegraphics[width=12.5cm,angle=0]{fig-12-top.eps}
\includegraphics[width=12.5cm,angle=0]{fig-12-bot.eps}
\caption{(top panels) Calculated and experimental kinematic moments
of inertia $J^{(1)}$ of the indicated one-quasiproton configurations
in the $^{241}$Am nucleus and ground state rotational band in
reference even-even $^{240}$Pu nucleus. Experimental data are shown in
the middle panel, while the results of the CRHB+LN calculations with the
NL1 and NL3* CEDF's in the left and right panels, respectively. 
The same symbols/lines are used for the same theoretical and experimental 
configurations. The symbols are used only for the configurations in odd-mass 
nucleus; the ground state rotational band in reference even-even nucleus 
is shown by solid black line. The label with the following structure
``Odd nucleus = reference even+even nucleus + proton($\pi$)/neutron($\nu$)''
is used in order to indicate the reference even-even nucleus and the type 
of the particle (proton or neutron) active in odd-mass nucleus. (bottom 
panels) The same as in top panels but for one-quasineutron configurations 
in $^{237}$U and ground state band in $^{236}$U. The experimental data are 
from Refs.\ \cite{Pu240,241Am-237Np}. Based on Ref.\ \cite{AO.13}.} 
\label{fig.ch8:12}
\end{figure}

\begin{figure}[h]
\centering
\includegraphics[width=12.6cm,angle=0]{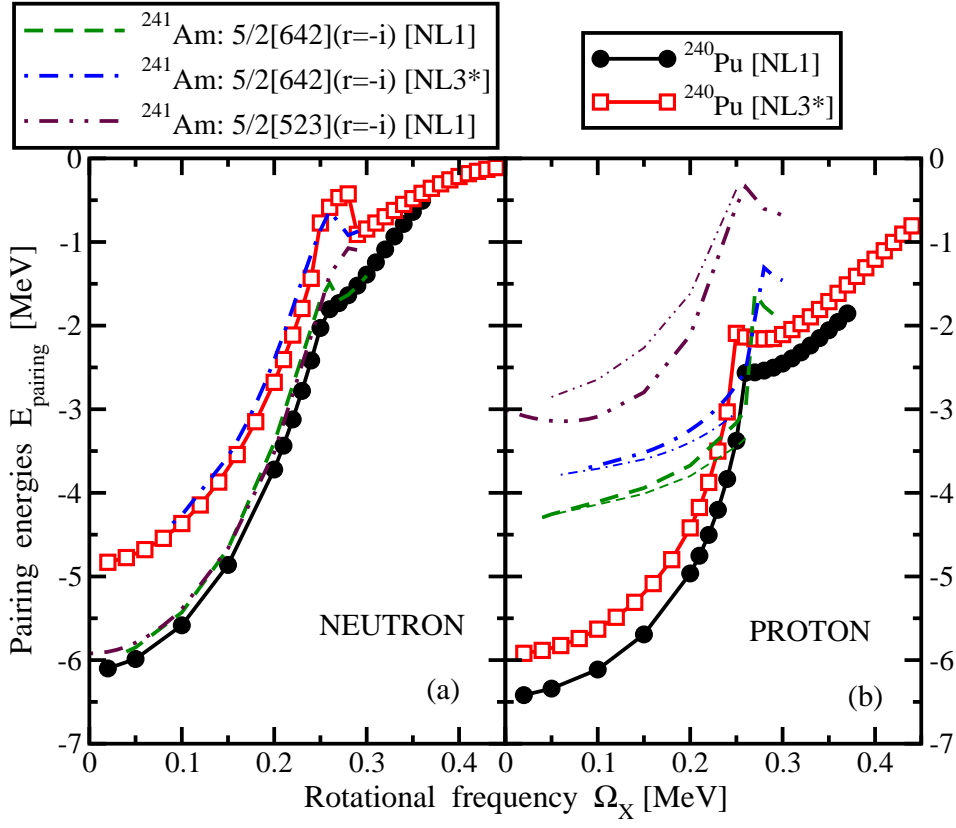}
\caption{Calculated proton and neutron pairing energies in ground 
state rotational band of $^{240}$Pu and one-quasiproton rotational 
bands of $^{241}$Am. Thick and thin lines are used for the $(r=-i)$ 
and $(r=+i)$ branches of one-quasiparticle configurations, respectively.  
Note that neutron pairing almost does not depend on the signature of 
blocked proton orbital. As a result, only the $(r=-i)$ branches 
are shown in panel (a). From Ref.\ \cite{AO.13}.} 
\label{fig.ch8:13}
\end{figure}

  In the DFT framework, the description of rotational bands in 
odd-mass nuclei is more technically difficult than the one in 
even-even nuclei because of the reasons discussed below. The
systematic study of such bands has so far been performed  only
in CDFT \cite{AO.13}.

  First, the effects of blocking due to odd particle have to be 
included in a fully self-consistent way. This is done in the CRHB+LN 
computer code according to Refs.\ \cite{Ring1970Z.Phys.A10,EMR.80,Ring1980}. In
addition, it requires the identification of blocked orbital at
all frequencies of interest which is non-trivial problem. In
the CRHB+LN code \cite{A250,AO.13} the blocked orbital can be 
specified by different fingerprints such as 
\begin{itemize}
\item
  dominant main oscillator quantum number $N$ of the wave
  function, 

\item
  the dominant $\Omega$ quantum number ($\Omega$ is the projection 
  of the angular momentum on the symmetry axis) of the wave 
  function,

\item
  the particle or hole nature of the blocked orbital,

\item
  the position of the state within specific 
  parity/signature/dominant $N$/dominant $\Omega$ block,

\end{itemize} 
or their combination. For a given configuration, possible 
combinations of the blocked orbital fingerprints were defined 
from the analysis of calculated quasiparticle spectra in 
neighboring even-even nuclei and the occupation probabilities
of the single-particle orbitals of interest in these nuclei.

  Second,  variational calculations with blocked orbital(s) are numerically
less stable than the ones for the ground state bands in even-even nuclei
because at each iteration of the variational procedure blocked orbital 
has to be properly identified. The convergence problems are the consequence
of the fact that closely lying orbitals within a given parity/signature 
block interact and exchange a character; the strength of the interaction
is important factor here. Another ingredient affecting the convergence 
is the relative energies of interacting orbitals. Different CEDF's are 
characterized by different single-particle spectra \cite{AS.11}. As a result, 
the convergence problems for specific blocked solution can show up in one 
functional but will not affect the solution in another one. A detailed
discussion of the convergence problems in the calculations of rotational
bands in odd-mass nuclei and the ways to overcome them is presented in
Sect. V of Ref.\ \cite{AO.13}.

  Fig.\ \ref{fig.ch8:12} displays a representative example of the RHB+LN 
calculations for one-quasiparticle bands in $^{237}$Np and $^{241}$Am 
which come from systematical studies of Ref.\ \cite{AO.13}. Additional 
results for $^{253}$No and $^{255}$Lr can be found in Refs.\ 
\cite{No253-EPJ,255Lr}\footnote{To my knowledge, in the DFT framework
two-quasiparticle rotational bands have been calculated only in a single 
case of $^{74}$Rb \cite{Rb74} and again in the CRHB+LN approach}. In $^{241}$Am, 
the rotational bands based on the Nilsson orbitals $\pi 5/2[642]$ (from the 
$i_{13/2}$ spherical subshell), $\pi 5/2[523]$ (from the $h_{9/2}$ subshell) 
and  $\pi 3/2[521]$ (from the $f_{7/2}$ subshell) have been observed;
their kinematic moments of inertia $J^{(1)}$ are distinctly different
at low frequencies. Theoretical calculations (Fig.\ \ref{fig.ch8:12}a,c) 
describe well the absolute values of the kinematic moments of inertia of 
different configurations, their  evolution with rotational frequency and 
signature splitting. They also indicate that the results of the RHB+LN 
calculations for a specific configuration only weakly depend on CEDF. 
Above mentioned features are also clearly seen in $^{237}$Np.

  Figs.\ \ref{fig.ch8:12}a and c also shows that the convergence depends
on CEDF. Indeed, it was not possible to get convergent solutions for 
the  $\pi 5/2[523]$ and $\pi 3/2[521]$ configurations of $^{241}$Am 
in the calculations with the NL3* CEDF.

  The increase of the kinematic moment of inertia in the bands of 
$^{241}$Am as compared with the one of the ground state band in $^{240}$Pu 
(Fig.\ \ref{fig.ch8:12}) is caused by the blocking effect which results 
in a decreased proton  pairing (see Fig.\ \ref{fig.ch8:13}). This figure
also shows that the blocking of proton orbitals almost does not affect 
the pairing in neutron subsystem.

  The systematic studies of Ref.\ \cite{AO.13} allowed to conclude that
rotational properties of one-quasiparticle configurations substantially
depend on the structure of blocked orbital. As a result, the rotational 
properties reflected through the following fingerprints
\begin{itemize}
\item the presence or absence of signature splitting,

\item the relative properties of different configurations with 
respect of each other and/or with respect to the ground state
band in  reference  even-even nucleus,

\item the absolute values of the kinematic moments of inertia 
(especially at low rotational frequencies) and their evolution 
with rotational frequency

\end{itemize}
provide useful tools for quasiparticle configuration assignments.
Such configuration assignments are important, for example, for 
on-going experimental investigations of odd-mass light superheavy 
nuclei at the edge of the region where spectroscopic studies 
are still feasible (the nuclei with masses $A\sim 255$ and 
proton  number $Z\geq 102$) \cite{HG.08,AO.13}.  
Ref.\ \cite{AO.13} clearly showed that with few exceptions these 
features of rotational bands are well described in the RHB+LN 
calculations. The presence or absence of signature separation 
and its magnitude is the most reliable fingerprint which is reproduced 
in model calculations with good accuracy. The moments of inertia and 
their evolution with frequency are generally well described in model 
calculations. As a consequence, the relative properties of different 
configurations with respect of each other and/or with respect to the 
ground state band in reference even-even nucleus provide a reasonably 
reliable fingerprint of configuration.
 
  However, it is necessary to recognize that the configuration 
assignment based on rotational properties has to be complemented 
by other independent methods and has to rely on sufficient experimental
data \cite{AO.13}. This is because such method of configuration assignment not 
always leads to a unique candidate configuration due to theoretical 
inaccuracies in the description of the moments of inertia.

  The ability to calculate odd-mass nuclei fully self-consistently 
with allowance of nuclear magnetism and breaking of Kramer's degeneracy
has also allowed to address the question of consistency of the definition 
of pairing  strength in CDFT. The strengths of pairing defined by means 
of the kinematic moments of inertia $J^{(1)}$ and three-point $\Delta^{(3)}$ 
indicators 
\begin{eqnarray}
\Delta ^{(3)}_{\nu}(N) = \frac{(-1)^N}{2} \left[ B(N-1) + B(N+1) - 2 B(N) \right],
\label{eq.ch8:23}
\end{eqnarray} 
defined from odd-even staggering of binding energies $B(N)$ (similar expression
holds also for proton  $\Delta_{\pi}^{(3)}$ indicator), strongly
correlate \cite{AO.13}. This is known result in non-selfconsistent models 
based on phenomenological Woods-Saxon or Nilsson potentials. 
However, this is non-trivial result in the DFT framework since 
time-odd mean fields (absent in phenomenological potentials) 
strongly affect the moments of inertia \cite{Afanasjev2010Phys.Rev.C34329} and have 
an impact on three-point $\Delta^{(3)}$ indicators \cite{Afanasjev2010Phys.Rev.C14309}. 
The definitions of pairing strength via these two observables 
are complimentary. This is because (i) it is difficult to disentangle proton 
and neutron contributions to pairing when considering the moments of 
inertia and (ii) the $\Delta^{(3)}$ indicators are affected by particle-vibration 
coupling and depend on correct reproduction of the ground 
states in odd-mass nuclei \cite{AO.13}.

\section{Hyperdeformation at high spin}

   The search for hyperdeformation (HD) at high spin still remains in the 
focus of attention of nuclear structure community. Although the attempts 
to observe discrete HD bands at high spin have not been succesful so far, 
there is a hope that next generation of detectors such as GRETA and AGATA 
will allow to observe such bands in the future.

  Already, some hints of the presence of hyperdeformation at high spin are
available. For example, Hyper-Long-HyperDeformed (HLHD) experiment at the 
EUROBALL-IV $\gamma$-detector array revealed some features expected 
for HD nuclei \cite{HD-exp-1,HD-exp-2,Hetal.06}. Although no discrete 
HD rotational bands have been identified, rotational 
patterns in the form of ridge-structures in three-dimensional (3D) rotational 
mapped spectra are identified with dynamic moments of inertia $J^{(2)}$ ranging 
from 71 to 111 MeV$^{-1}$ in 12 different nuclei selected by charged particle- 
and/or $\gamma$-gating. The four nuclei, $^{118}$Te, 
$^{124}$Cs, $^{125}$Cs and $^{124}$Xe, found with moment of inertia $J^{(2)}\sim 110$
MeV$^{-1}$ are most likely hyperdeformed \footnote{In comparison, the HD ridges 
in $^{152}$Dy are characterized by $J^{(2)}\sim 130$ MeV$^{-1}$ \cite{152Dy-exp-2}.} while 
the remaining nuclei with smaller values of $J^{(2)}$ are expected to be 
superdeformed. The width in energy of the observed ridges indicates that there 
are $\approx 6-10$ transitions in the HD cascades, and a fluctuation analysis 
shows that the number of bands in the ridges exceeds 10. The HD ridges are
observed in the frequency range of about 650 to 800 keV, and their dynamic 
moments of inertia have typical uncertainty of 10\% (e.g. $111\pm 11$ MeV$^{-1}$ 
in $^{124}$Xe).

\begin{figure}[ht]
\centering
\vspace{1.5cm}
\includegraphics[angle=0,width=12.0cm]{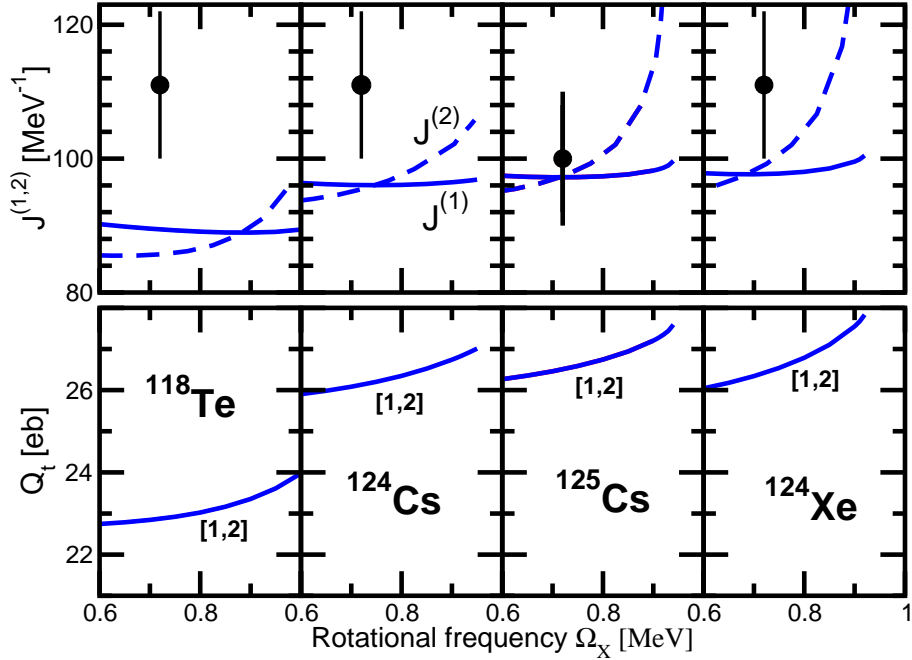}
\caption{Calculated kinematic and dynamic moments of inertia 
(top panels) and transition quadrupole moments (bottom panels) 
as a function of rotational frequency for the lowest HD solutions 
in $^{118}$Te, $^{124,125}$Cs and $^{124}$Xe. The structure of calculated 
configurations is indicated at bottom panels. Experimental data for 
dynamic moments of inertia of ridge structures are shown in top 
panels. From Ref.\ \protect\cite{AA.08}.}
\label{fig.ch8:14}
\end{figure}

\begin{figure}[ht]
\centering
\vspace{0.5cm}
\includegraphics[angle=0,width=12.0cm]{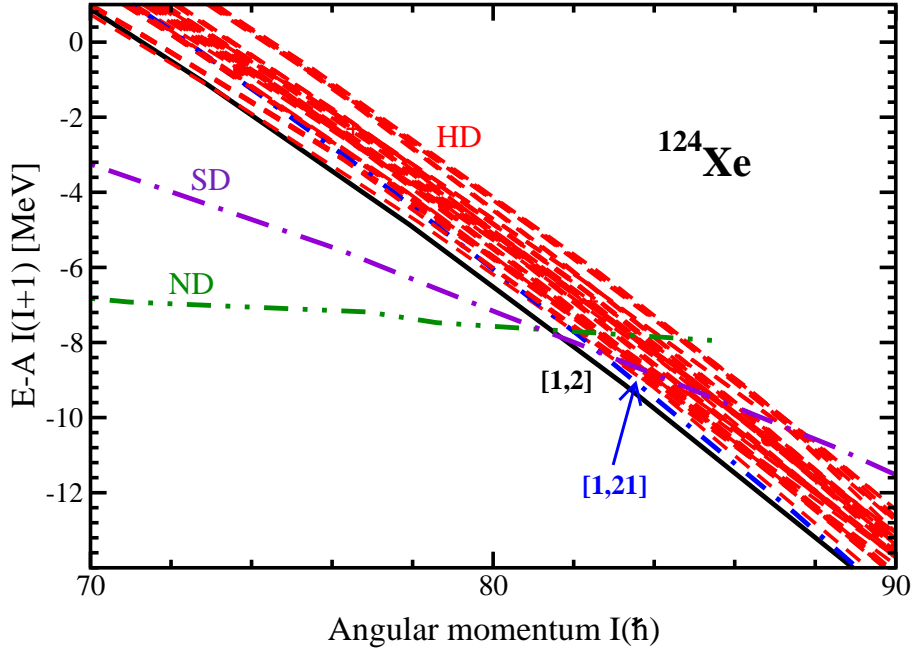}
\caption{Energies of the calculated configurations relative to a smooth liquid 
drop reference $AI(I+1)$, with the inertia parameter $A=0.01$. The ND and SD 
yrast lines are shown by dotted and dot-dot-dashed lines, respectively. Solid 
and dot-dashed lines are used for the [1,2] and [1,21] HD configurations, 
respectively. Dashed lines represent excited HD configurations. From Ref.\ 
\protect\cite{AA.08}.}
\label{fig.ch8:15}
\end{figure}

  The experimental data show unusual features never seen before in the studies 
of the SD bands.  For example, the addition of one neutron on going from 
$^{124}$Cs to $^{125}$Cs decreases the experimental $J^{(2)}$ value by $\sim 10\%$ 
(from 111 MeV$^{-1}$ down to 100 MeV$^{-1}$, see Fig.\ \ref{fig.ch8:14}). 
It is impossible to find an explanation for such a big impact of the single 
particle on the properties of nuclei: previous studies in the SD minima in 
different parts of the nuclear chart showed that the addition or removal of particle 
affects dynamic moments of inertia less drastically \cite{ALR.98}.

  The comparative analysis of the CRMF and CRHB+LN results in $^{124}$Xe
in Ref.\  \cite{AA.08} reveals that the pairing is reasonably small in 
the lowest HD configurations and it becomes even smaller in excited
configurations due to blocking effect. The dominant effects in the 
quenching of pairing correlations are the Coriolis antipairing effect, 
the quenching due to shell gaps \cite{SGBGV.89}, and the blocking  effect 
\cite{Ring1980}.  Thus, the pairing has been neglected in systematic 
studies of HD in Ref.\ \cite{AA.08} in which unpaired 
CRMF calculations have been performed.

\begin{figure}[ht]
\centering
\includegraphics[angle=0,width=3.2cm]{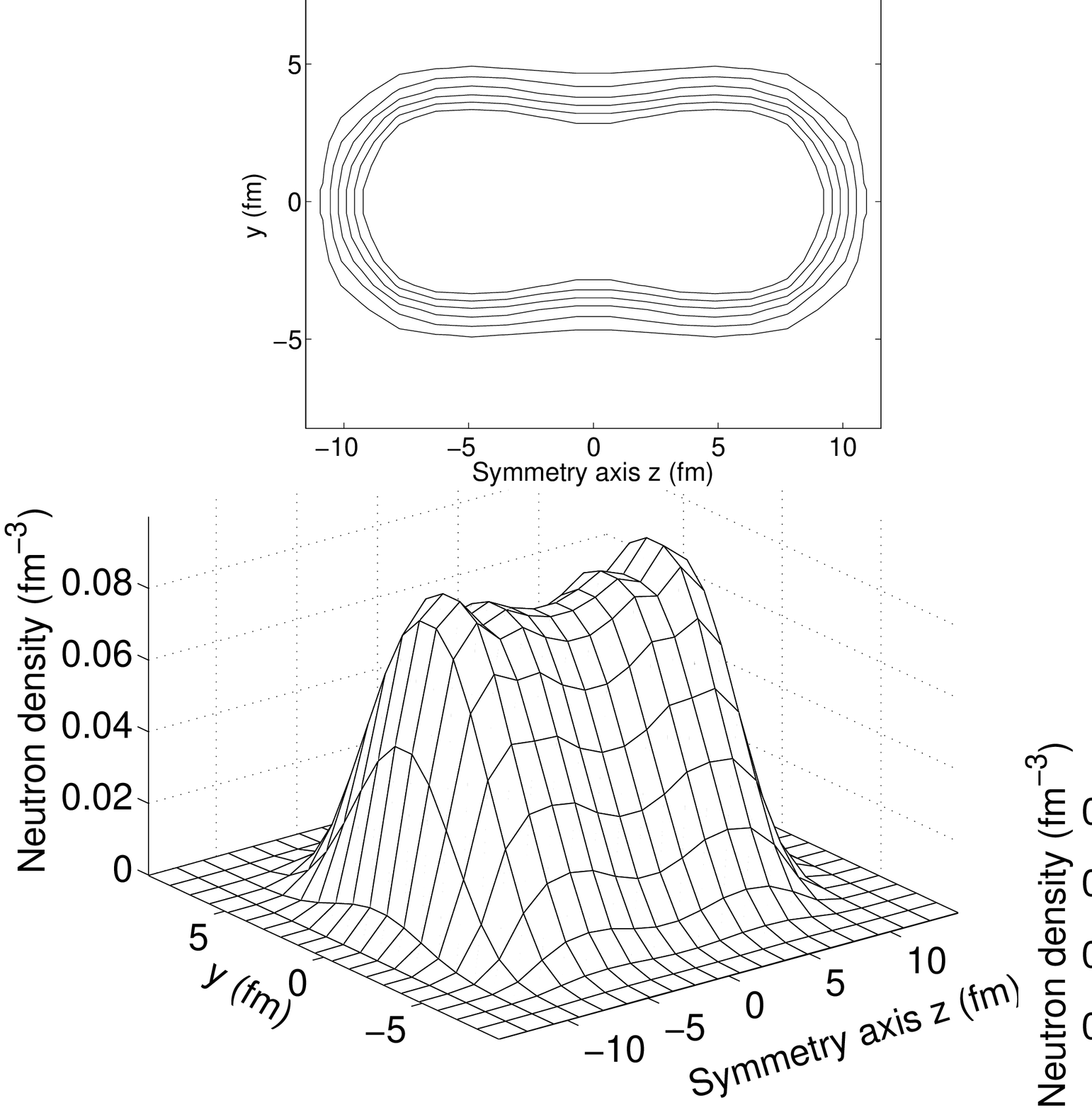}
\caption{(left panels) The self-consistent neutron density $\rho_n(y,z)$ 
as a function of $y$- and $z$- coordinates for the [1,2] configuration 
in $^{124}$Xe at rotational frequency $\Omega_x=0.75$  MeV. Top and bottom 
panels show 2- and 3-dimensional plots of the density distribution, 
respectively. In the top panel, the densities are shown in steps of 0.01 
fm$^{-3}$ starting from $\rho_n(y,z)=0.01$ fm$^{-3}$. (right panels)
The same as left panels, but for yrast megadeformed state in $^{102}$Pd 
at rotational frequency $\Omega_x=0.95$ MeV. Two top panels show 2-dimensional 
plots of the proton and neutron densitydistribution. From Ref.\ \cite{AA.08}.}
\label{fig.ch8:16}
\vspace{12.5cm}
\end{figure}
 
  The results of the CRMF calculations with the NL1 CEDF are 
confronted with experimental data in Fig.\ \ref{fig.ch8:14}.  The calculated $J^{(2)}$ 
moments of inertia somewhat underestimate experimental data. The kinematic moments 
of inertia of the lowest HD solutions are either nearly constant or very gradually 
increase with rotational frequency. On the contrary, both the dynamic moments of 
inertia and  the transition quadrupole moment $Q_t$ more rapidly increase with
rotational frequency over the calculated frequency range. They are in complete 
contract to the features of the SD bands in unpaired regime, in which the $Q_t$, 
$J^{(1)}$ and $J^{(2)}$ values (apart from the unpaired band crossing regions) 
decrease with increasing rotational frequency (see Refs.\ 
\cite{BRA.88,AKR.96,A60,Vretenar2005Phys.Rep.101}  and references therein). As discussed in detail in 
Ref.\ \cite{AA.08} the microscopic origin of these features lies in a centrifugal 
stretching of the HD shapes with increasing rotational frequency. Systematic analysis 
of the yrast/near-yrast HD configurations in the part of the nuclear chart studied in 
Ref.\ \cite{AA.08} shows that the centrifugal stretching is a general feature of the 
HD bands.

  In order to better understand the general features of HD at high spin I
concentrate on the $^{124}$Xe nucleus. The results of the CRMF calculations 
for some HD configurations in $^{124}$Xe are displayed in Fig.\ \ref{fig.ch8:15}.
The calculated configurations are labeled by $[p,n_1n_2]$, where $p$, $n_1$ and 
$n_2$ are the number of proton $N=7$ and neutron $N=7$ and $N=8$ hyperintruder 
orbitals occupied, respectively.  For most of the HD configurations, neutron 
$N=8$ orbitals are not occupied, so the label ${n_2}$ is omitted in the labeling 
of such configurations. The HD minimum becomes lowest in energy at spin $82\hbar$, and the [1,2]
configuration is the yrast HD configuration in the spin range of 
interest. The excited HD configurations displayed in Fig.\ \ref{fig.ch8:15} 
are built from this configuration by exciting either one proton or 
one neutron or simultaneously one proton and one neutron. The total 
number of excited HD configurations shown is 35. It is interesting to 
mention that the configuration involving the lowest $N=8$ neutron 
orbital (the [1,21] conf. in Fig.\ \ref{fig.ch8:15}) is calculated 
at low excitation energy.

  The calculations reveal a high density of the HD configurations which will be 
even higher if the additional calculations for the excited configurations would 
be performed starting from the low-lying excited HD configurations, such as the 
[1,21] configuration. This high density is due to two facts: relatively small 
$Z=54$ and $N=70$ HD shell gaps in the frequency range of interest (see Fig.\ 
10 in Ref.\ \cite{AA.08} and the softness of the potential energy surfaces in 
the HD minimum.

  With few exceptions mentioned below these two factors are active at 
hyperdeformation in absolute majority of even-even nuclei in the $Z=40-60$ 
part of nuclear chart studied in Ref.\ \cite{AA.08}. Note that different 
proton and neutron HD gaps will be active in different regions \cite{AA.08}. 
As a consequence, the density of the HD bands in the spin range where they 
are yrast or close to yrast is high in the majority of the cases. For such 
densities, the feeding intensity of an individual HD band will most likely 
drop below the observational limit of modern experimental facilities so that 
it will be difficult to observe discrete HD bands. This is because total 
feeding intensity will be redistributed among many low-lying bands, 
thus drastically reducing the intensity with which each individual band is populated. 
On the other hand, the high density of the HD bands may favor the observation of 
the rotational patterns in the form of ridge-structures in three-dimensional 
rotational mapped spectra as it has been seen in the HLHD experiment \cite{Hetal.06}. 
The study of these patterns as a function of proton and neutron numbers, which 
seems to be possible with existing facilities, will provide a valuable information 
about hyperdeformation at high spin.

\begin{figure}[ht]
\centering
\includegraphics[angle=0,width=6.5cm]{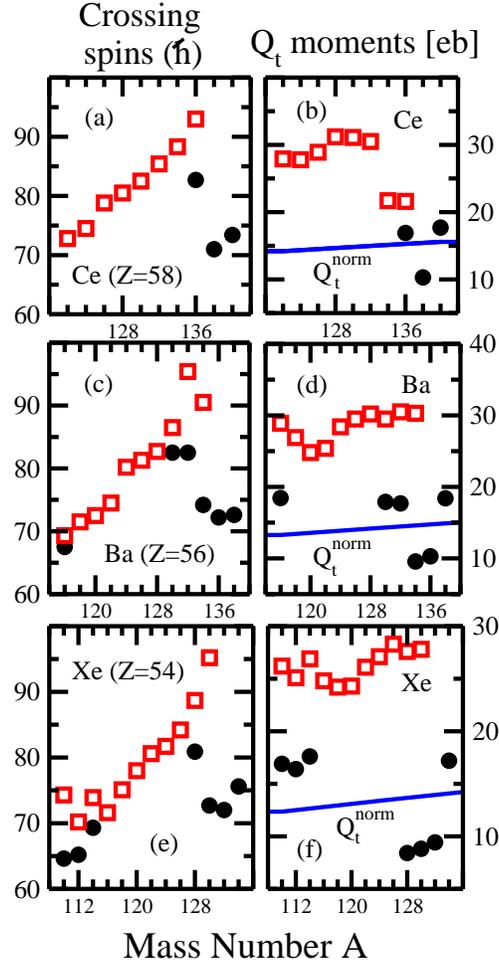}
\vspace{0.4cm}
\caption{The crossing spins (left panels) at which the SD 
(solid circles) and HD (open squares) configurations become yrast and 
their transition quadrupole moments $Q_t$ (right panels) for the Ce, 
Ba and Xe isotopes. The values for the SD configurations are shown only 
in the cases when they become yrast at lower spins than the HD configurations. 
The normalized transition quadrupole moments $Q^{norm}_t$ corresponding 
to the deformation of the yrast SD band in $^{152}$Dy are shown by blue
solid line. Based on Fig. 4 from Ref.\ \cite{AA.08}.}
\label{fig.ch8:17}
\end{figure}

 Only in few nuclei the density of the HD rotational bands is low in the CRMF 
calculations \cite{AA.08}. These are, for example, $^{96}$Cd and $^{107}$Cd 
\cite{Afanasjev-PhysRevC.72.031301-2005,Abusara-PhysRevC.79.024317-2009}.
The later one is the best candidates for a search of discrete HD bands.
An alternative candidate is the doubly magic extremely superdeformed 
band in $^{111}$I \cite{AA.08}, the deformation of which is only slightly 
lower than that of the HD bands, and which may be observed with existing 
experimental facilities. In all these cases the low density of the HD
bands is due to large shell gaps in the single-particle spectra at HD.

  Another interesting question is whether necking degree of freedom is 
important in the HD bands. Fig.\ \ref{fig.ch8:16} shows some indications of 
the necking and the clusterization of the density into two fragments in the 
[1,2] configuration of $^{124}$Xe, but this effect is not very pronounced in 
this nucleus. The systematics of the self-consistent proton density distributions 
in the HD states has been studied in Ref.\ \cite{AA.08}. In some nuclei (such as 
$^{124}$Te, $^{130}$Xe, $^{132}$Ba) the necking degree of freedom plays an important 
role, while others (for example, $^{100}$Mo and $^{136}$Ce) show no necking. The neck 
is typically less pronounced in the HD states of the lighter nuclei because of their 
smaller deformation.  In addition, the shell structure also plays a role in a 
formation of neck. The necking degree of freedom becomes even more important in 
extremely deformed structures which according to the language of Ref.\ \cite{DPSD.04} 
can be described as megadeformed. Fig.\ \ref{fig.ch8:16} shows an example of density 
distribution for the megadeformed state in $^{102}$Pd, which becomes yrast at 
$I\sim 85\hbar$ in the CRMF calculations. The neck is more pronounced in the proton 
subsystem  than in the neutron one both in the HD and megadeformed structures  due 
to the Coulomb repulsion of the segments. These results indicate that, in general, 
the necking degree of freedom is important in the HD states and that it should be 
treated within the self-consistent approach which, in particular, allows different 
necking for the proton and neutron  subsystems.

 The spins $I^{HD}_{cr}$ at which HD bands become yrast (or shortly ``crossing spins'') 
represent an important constraint for observation of such bands. There are several 
reasons for that. First, the sensitivity of modern $\gamma$-ray detectors defines 
the highest spins which can be studied in experiment. For example, the GAMMASPHERE
allows to study discrete rotational bands only up to $\approx 65\hbar$ 
in medium mass nuclei 
\footnote{The DFT calculations suggest that the triaxial SD band 1 in $^{158}$Er 
is observed up to spin in excess of 70$\hbar$ \cite{ASN.12}. However, 
this result requires further experimental confirmation.}.  The observation 
of higher spin states will most likely require a next generation of 
$\gamma$-ray tracking detectors such as GRETA or AGATA. Second,  as suggested by the 
studies of the Jacobi shape transition in  Ref.\ \cite{SDH.07},  the coexistence of 
the SD and HD minima at the feeding spins may have an impact on the survival of the 
HD minima because of the decay from the HD to  SD configurations. If this mechanism is 
active,  then only the nuclei in which the HD minimum is lower in energy than the 
SD one at the feeding spin and/or the nuclei characterized by the large 
barrier between the HD and SD minima will be the reasonable candidates 
for a search of the HD bands.

\begin{figure}[ptb]
\centering
\includegraphics[angle=0,width=16.3cm]{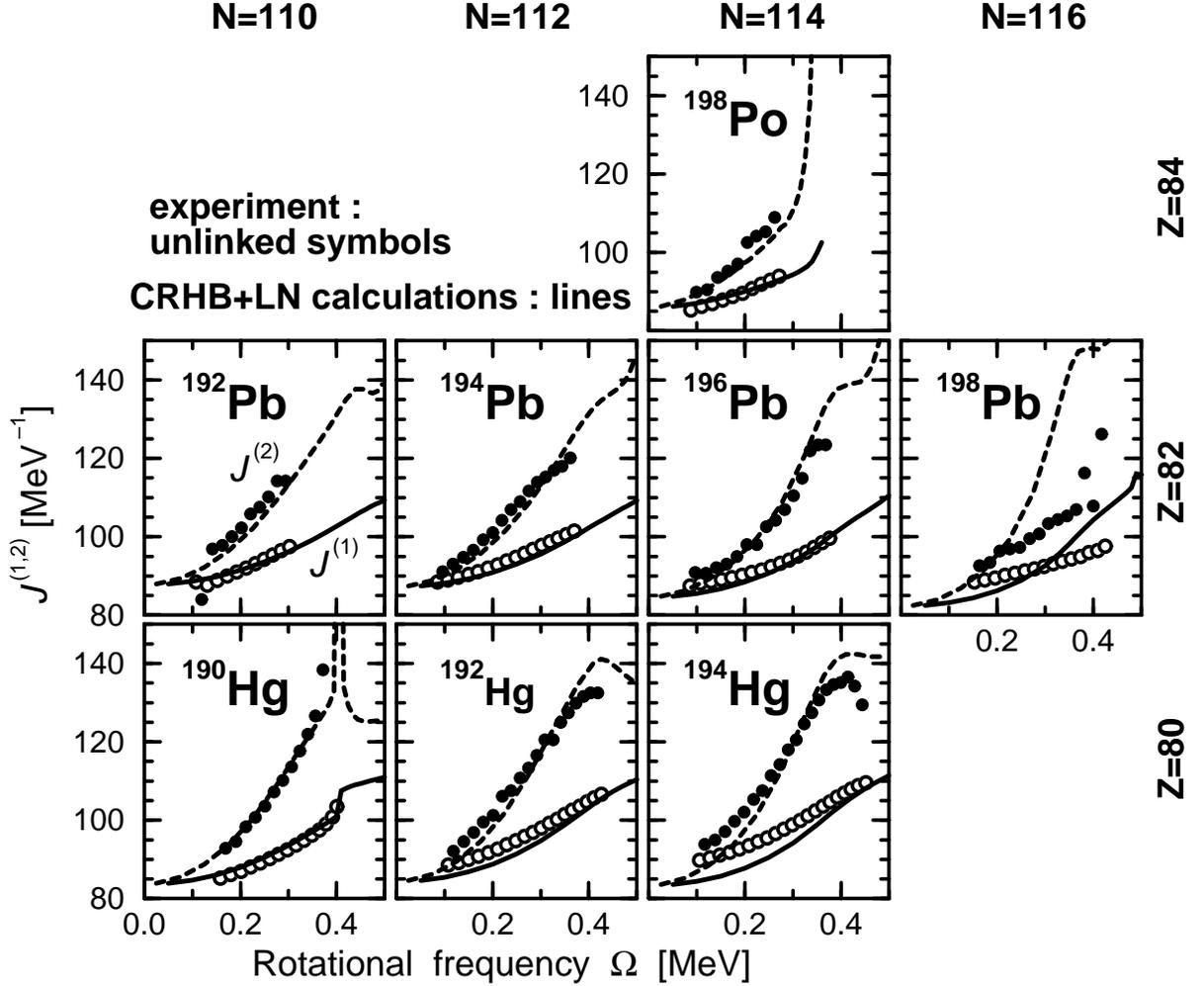}
\caption{Kinematic ($\mathcal{J}^{(1)}$) and dynamic
($\mathcal{J}^{(2)}$) moments of inertia of the yrast
superdeformed bands in the even-even nuclei of the
$A\sim190$ mass region. Experimental $\mathcal{J}^{(1)}$ 
and $\mathcal{J}^{(2)}$ values are shown by open and solid 
circles. Solid and dashed lines are used for the 
$\mathcal{J}^{(1)}$ and $\mathcal{J}^{(2)}$ values obtained in the
CRHB+LN calculations. From Ref.~\protect\cite{Afanasjev2000Nucl.Phys.A196}.}
\label{fig.ch8:18}
\end{figure}

   Figs.\ \ref{fig.ch8:17} shows that the HD configurations become  yrast at 
lower spin than the SD ones only in a specific mass range which depends on the 
isotope chain \cite{AA.08}. It also illustrates that the crossing spins $I_{cr}^{HD}$, 
at which the HD configurations become yrast, are lower for proton-rich nuclei. 
This is a feature seen in the most of studied isotope chains; by going from the 
$\beta$-stability valley towards the proton-drip line one can lower $I_{cr}^{HD}$
by approximately $10\hbar$. In addition, the calculated transition quadrupole 
moments of these configurations at spin values close to the crossing spins are 
shown. The calculated HD configurations are near-prolate. Additional results 
for the Zr, Mo, Ru, Pd, Cd, Sn, Te isotope chains are presented in Ref.\ 
\cite{AA.08}.

 In addition, the single-particle properties and their role in future
configuration assignments at HD have been investigated in Ref.\ \cite{AA.08}.
It was concluded that the individual properties of the single-particle 
orbitals are not lost at HD. In the future, they will allow the assignment 
of the configurations to the HD bands using the relative properties of 
different bands. Such methods of configuration assignment were originally 
developed for superdeformation. In contrast to the case of SD, the analysis
of Ref.\ \cite{AA.08} has showed that only simultaneous application (by 
comparing experimental and theoretical $(i_{eff},\Delta Q_t)$ values)
of the methods based on effective alignments $i_{eff}$ \cite{Rag.93} and 
relative transition quadrupole moments $\Delta Q_t$ \cite{SDDN.96}  will 
lead to a reliable configuration assignment for the HD bands. 
Moreover, additional information on the  structure of the HD bands will be 
obtained from the band crossing features; the cases of strong interaction of 
the bands in unpaired regime at HD will be more common as compared with 
the situation at SD.

\section{Other phenomena}
\label{other-phenomena}
  
 It is impossible to cover in detail other results obtained for rotational 
bands within the cranked versions of the CDFT because of space limitations. 
However, I feel that at least it is necessary to give some short overview of 
systematic calculations in different mass regions and provide some references 
on the physical phenomena which have not been discussed in detail in this review. 
Note, that transition quadrupole moments are typically described within the 
error bars of experimental measurements 
\cite{AKR.96,Afanasjev1999Phys.Rev.C51303,Afanasjev2000Nucl.Phys.A196,AF.05,A.12}. 
Thus, we do not consider them below.

\subsection{Superdeformation in paired and unpaired regimes}

 Results of systematic CRHB+LN calculations with the NL1 CEDF 
\cite{Afanasjev1999Phys.Rev.C51303,Afanasjev2000Nucl.Phys.A196} 
for yrast superdeformed bands in even-even 
nuclei in the $A\sim 190$ region are shown in Fig.\ \ref{fig.ch8:18}. 
Without any new adjustable parameter a very successful description 
of rotational features of experimental bands in this region is 
obtained in the calculations;  only in $^{198}$Pb the calculations fail 
to reproduce the kinematic and the dynamic moments of inertia above 
$\Omega_x\sim0.25$ MeV (Fig.\ \ref{fig.ch8:18}). The increase of the moments of inertia 
in this mass region can be understood as emerging
predominantly from a combination of three effects: the gradual
alignment of a pair of $j_{15/2}$ neutrons, the alignment of a pair
of $i_{13/2}$ protons at a somewhat higher frequency, and decreasing
pairing correlations with increasing rotational frequency. Above
$\Omega_x\geq 0.4$ MeV, the $\mathcal{J}^{(2)}$ values determined by
the alignment in the neutron subsystem decrease but this process is
compensated by the increase of $\mathcal{J}^{(2)}$ due to the
continuing alignment of the $i_{13/2}$ proton pair. Thus the shape
of the peak (or the plateau) in the total value of $\mathcal{J}^{(2)}$ 
at these frequencies is determined by a delicate balance between 
alignments in the proton and neutron subsystems which depends on the 
deformation, the rotational frequency and the position of the 
Fermi energy. For example, the alignment of a pair of $j_{15/2}$ neutrons 
at $\Omega_x\sim 0.4$ MeV is visible experimentally in the isotope 
$^{192}$Hg and even more clearly in $^{194}$Hg.

\begin{figure}[ptb]
\centering
\includegraphics[angle=0,width=17.3cm]{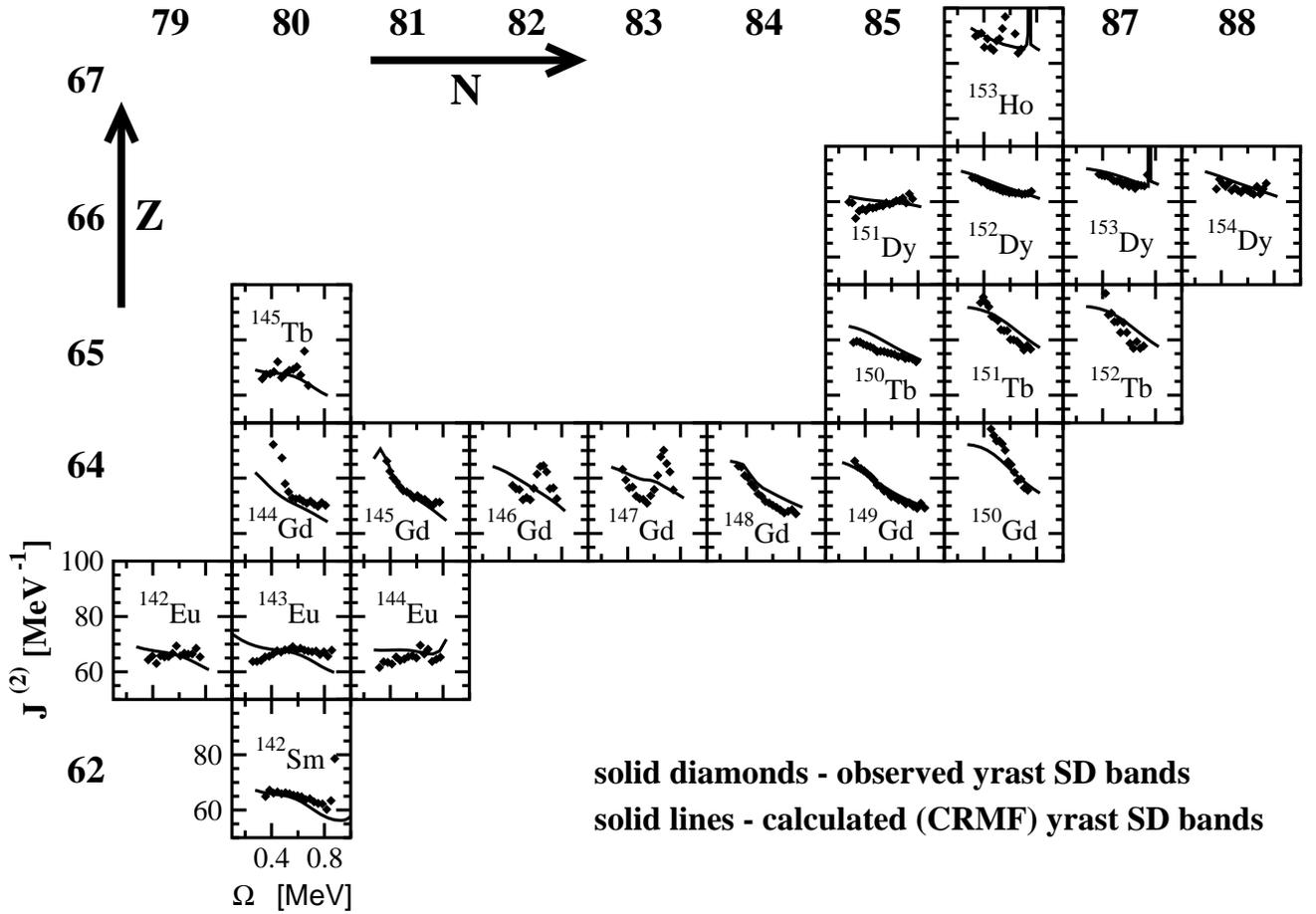}
\caption{Dynamic moments of inertia of observed yrast SD bands
(solid diamonds) in the $A\sim140-150$ mass region of
superdeformation compared with the calculations in cranked RMF
theory (solid lines). From Ref.\ \cite{Vretenar2005Phys.Rep.101}.}
\label{fig.ch8:19}
\end{figure}

   In addition to actinides, the CRHB+LN approach has been used for
the yrast SD band in $^{60}$Zn in Ref.\ \cite{Pingst-A30-60}.
 
\begin{figure}[th]
\centering
\includegraphics[angle=0,width=18.3cm]{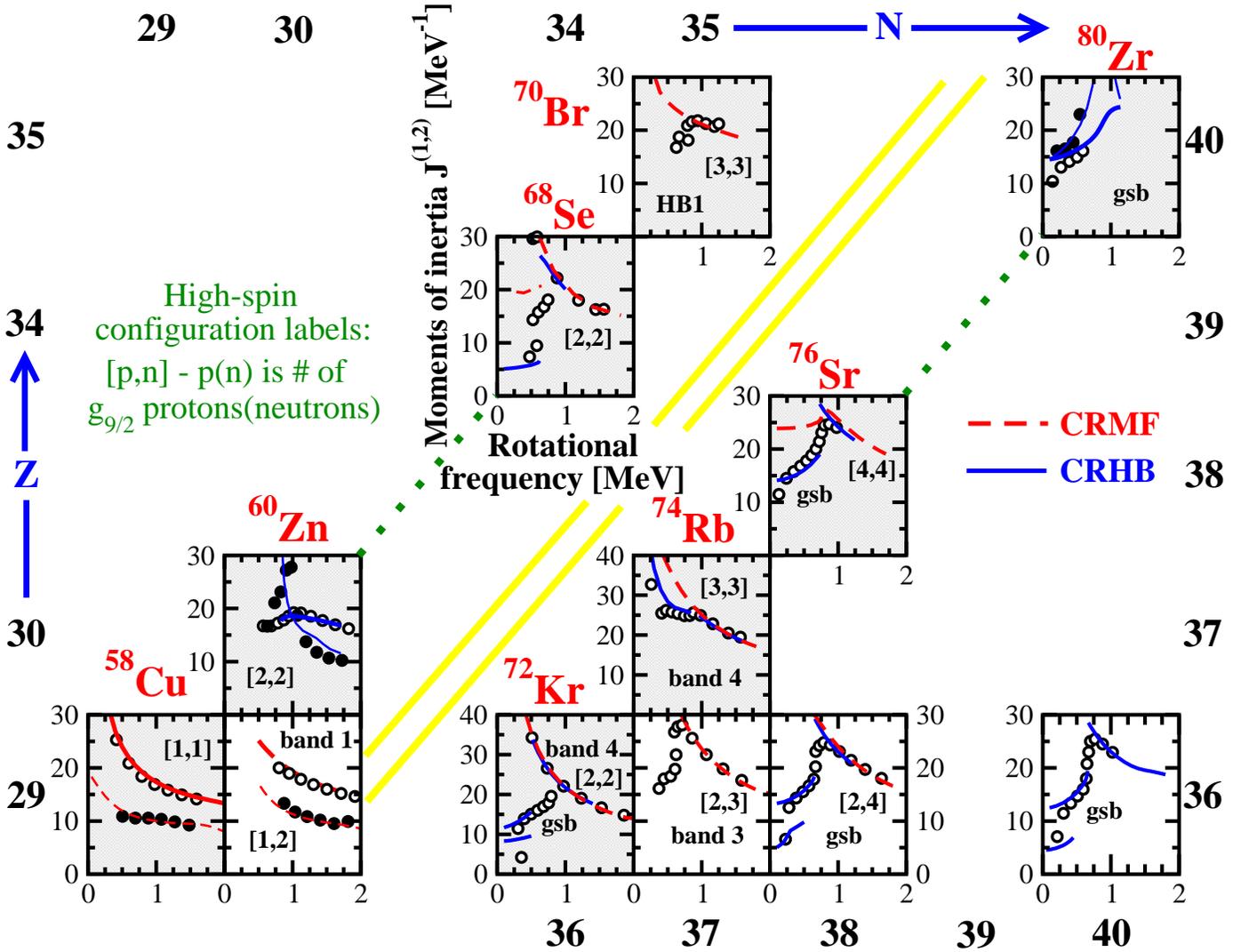}
\caption{The kinematic moments of inertia $J^{(1)}$ of rotational structures 
in the $N\approx Z$ nuclei compared with the results of the CRMF and 
CRHB+LN calculations. The shaded background is used for $N=Z$ nuclei. The 
vertical scale of the panels for $^{72}$Kr and $^{74}$Rb is different 
from the one of the other panels. Note that in few cases the results for 
dynamic moments of inertia $J^{(2)}$ are shown. In these cases, thick and thin 
lines are used for calculated kinematic and dynamic moments of inertia, 
respectively. Experimental kinematic and dynamic moments of inertia are shown 
by open and solid circles, respectively. The results of the CRHB+LN calculations 
at low spin are shown both for prolate and oblate minima in few cases; in a 
given nucleus calculated $J^{(1)}$ in oblate minimum is lower than the one in prolate minimum.
\label{fig.ch8:20}}
\end{figure}

 A systematic investigation of properties of superdeformed bands in
unpaired regime has been performed in the CRMF framework in the 
$A \sim 60$ \cite{A60} and 150 \cite{AKR.96,ALR.98} mass regions of 
superdeformation. It was shown that CRMF theory reproduces in general 
well the experimentally observed features.  For example, a summary of 
these studies for the dynamic moments of inertia $\mathcal{J}^{(2)}$ in 
the $A\sim 150$ region is shown in Fig.\ \ref{fig.ch8:19} and the left corner
of Fig.\ \ref{fig.ch8:20} shows the results for the $A\sim 60$ region.
I will discuss here the results for $A\sim 150$ mass region. At high 
rotational frequencies, where pairing is of minor importance, the dynamic 
moments of inertia are rather well reproduced in the CRMF calculations. 
Exceptions are the yrast superdeformed bands in $^{146,147}$Gd and in $^{153}$Ho 
undergoing unpaired band crossings. For example, in $^{146,147}$ Gd the peak in
$\mathcal{J}^{(2)}$ at $\Omega_x\approx 0.6$ MeV is not reproduced.
According to the standard interpretation \cite{Rag.93,ALR.98}, it
originates from the crossing of two specific orbitals with the
Nilsson quantum \ numbers $\nu \lbrack 651]1/2$ and $\nu \lbrack
642]5/2$ and signature $r=+i$. The relative position of this pair is
not reproduced properly in several RMF parameterizations
\cite{AKR.96,ALR.98}. Even at low spins, the results of the CRMF
calculations are close to experiment in most of the cases. The experiment 
of Ref.\ \cite{LCD.02} has linked the yrast superdeformed band
in $^{152}$Dy to the low-spin level scheme which allowed to extract
experimental kinematic moment of inertia. Its calculated value is by 
7-5\% higher than in experiment at low spins and the difference decreases 
with increasing spin (Ref.\ \cite{Vretenar2005Phys.Rep.101}. This suggests some persistence 
of pairing correlations especially at low spins. Note that the systematics
of Fig.\ \ref{fig.ch8:19} has recently been extended by the CRMF studies
of excited SD bands in $^{154}$Dy \cite{154Dy}.

\subsection{Neutron-proton pairing}

  The physics of isoscalar and isovector neutron-proton pairing has 
been and still is an active topic of nuclear structure studies 
\cite{A.12,FM.14}. At present, the existence of isovector $np$-pairing is 
well established. The isovector $np-$pairing is absolutely necessary in 
order to restore the isospin symmetry of the total wave function. Its strength is well 
defined by the isospin symmetry. On the contrary, the observed consequences of 
the $t=0$ $np-$pairing still remain illusive.  The systematic analysis of the 
rotational response of  $N\approx Z$ nuclei performed in the CRHB+LN and CRMF
frameworks (see Refs.\ \cite{AF.05,A.12} and references quoted therein) agrees with the 
picture which does not involve isoscalar $np-$pairing. According to it (isovector 
mean-field theory \cite{FS.99-NP}), at low spin, an isoscalar $np-$pair field is absent 
while a strong isovector pair field exists, which includes a large $np$ component, 
whose strength is determined by isospin conservation. Like in nuclei away from 
the $N = Z$  line, this isovector pair field is destroyed by rotation. In this 
high-spin regime, calculations without pairing describe accurately the data
(see Fig.\ \ref{fig.ch8:20}  and Refs.\ \cite{AF.05,A.12}), provided that the shape 
changes and band termination are taken into account.

\subsection{Band termination}

\begin{figure}[ht]
\centering
\includegraphics[angle=0,width=13.3cm]{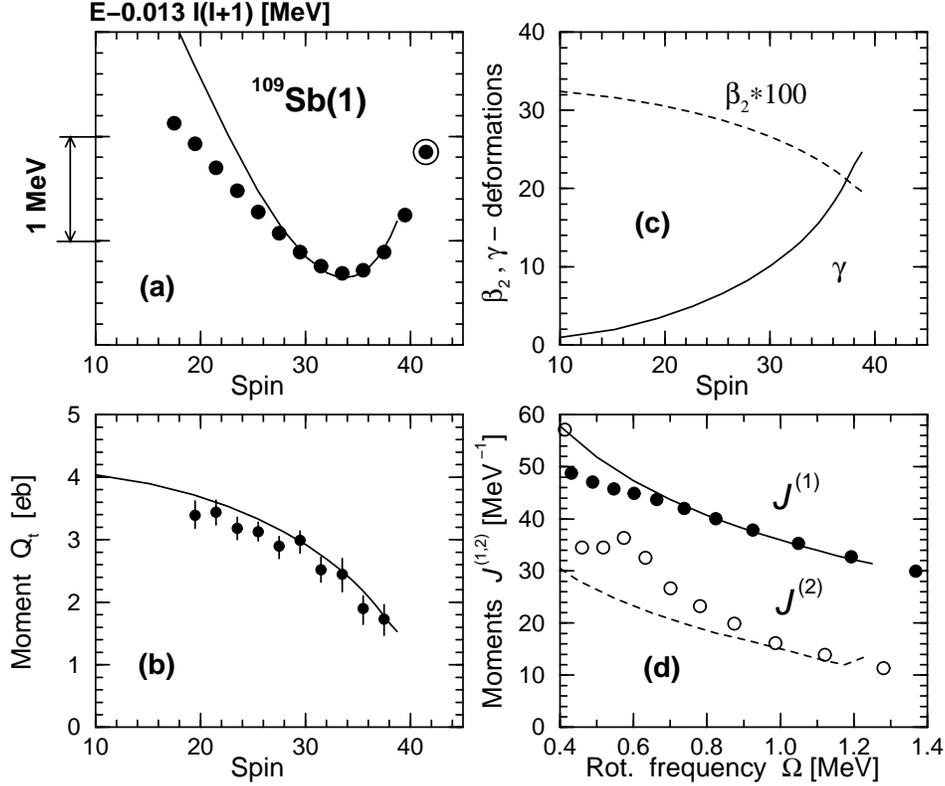}
\caption{Smooth terminating band 1 in $^{109}$Sb. Circles indicate
experimental data, while the results of CRMF calculations are
shown by lines. Panel (a) shows the excitation energies relative
to a rigid rotor reference $E_{RLD}(I)=0.013I(I+1)$ MeV. The
terminating state is indicated by a large open circle. The results
of the calculations are normalized to experiment at the position
of the minima in the ($E-E_{RLD}$) curve. Panel (b) compares
measured and calculated transition quadrupole moments $Q_{t}$.
Calculated $\protect\beta_{2}$ and $\protect\gamma$ deformations
are shown in panel (c). Panel (d) shows the calculated and
experimental kinematic and dynamic moments of inertia.
From Ref.\ \cite{Vretenar2005Phys.Rep.101}.}
\label{fig.ch8:21}
\end{figure}

  Another interesting phenomenon is band termination, and, especially 
smooth band termination. Smooth terminating band shows a continuous 
and smooth transition within a specific  configuration from a collective 
rotation to a non-collective terminating (single-particle) state \cite{Afanasjev1999Phys.Rep.1}. 
The band terminates in the terminating state which shows full alignment 
of all angular momentum vectors of the valence particles and holes along
the axis of rotation. In the CDFT framework, the band termination phenomenon
has been studied only in a few cases. One of them is ground state band in 
$^{20}$Ne terminating at $I=8^+$; this is a classical case of band termination.
The termination of this band and the impact of time-odd mean fields on its
rotational properties and terminating state have been studied in Ref.\ 
\cite{A.08}. In addition, the impact of these fields on the binding energies 
of terminating states in the $A\sim 44$ region has been studied in Ref.\ 
\cite{A.08}. Fig.\ \ref{fig.ch8:21} presents the only case of the study of smooth
band termination in the CRMF framework; this is smooth terminating band 1 in 
$^{109}$Sb.
One can see that experimental observables shown in Fig.\ \ref{fig.ch8:21}
are well described in the CRMF calculations 
at $\Omega_x \geq 0.8$ MeV (above $I\geq 25~\hbar$) where pairing is expected 
to play a minor role. The discrepancies between experiment and theory seen 
at lower spins are probably due to neglect of the pairing correlations in the 
CRMF calculations. One should note that without special techniques
(as developed in cranked Nilsson-Strutinsky approach \cite{Afanasjev1999Phys.Rep.1}) it is
impossible to trace this configuration in the CRMF calculations up to its
terminating state.
 
  It is necessary to recognize that not all rotational bands will terminate
in non-collective terminating state at the maximum spin $I_{max}$
\cite{Afanasjev1999Phys.Rep.1}. So far the examples of non-termination of 
rotational bands at the maximum spin have been experimentally observed and 
studied in the CRMF framework only in $^{74}$Kr and $^{75}$Rb \cite{74Kr,75Rb}.
  
\subsection{Other results}

  Other phenomena in rotating nuclei or the features of rotational bands
have been studied in the CRMF/CRHB frameworks during last decade. They 
will only be briefly mentioned here.

  The additivity principle for quadrupole moments and relative alignments has been
studied on the example of highly-deformed  and superdeformed bands in the
$A\sim 130$ mass region in Ref.\ \cite{MADLN.07}.   This principle of 
the extreme shell model stipulates that an average value of a one-body operator
be equal to the sum of the core contribution and effective contributions of 
valence (particle or hole) nucleons. It is only valid in an unpaired regime typical 
of high angular momenta since the pairing smears out the individuality of each 
single-particle orbital. The additivity principle for angular momentum alignments does 
not work as precisely as it does for quadrupole moments.

  Triaxial superdeformed (TSD) rotational bands at ultra-high spin in the Er region have
been studied on the example of $^{158}$Er in Ref.\ \cite{ASN.12}. Good description of
the deformation and rotational properties of the TSD bands observed in this nucleus has
been achieved in the CRMF calculations. Based on the results of covariant and 
non-relativistic DFT calculations it was suggested  that the TSD band 1 in this nucleus 
is observed up to spin in excess of 70$\hbar$ \cite{ASN.12} which is the highest spin 
reported so far. The enhanced deformation and SD bands in the Hf isotopes have been investigated 
in Ref.\ \cite{171-172Hf}. Contrary  to previous claims of triaxiality of the SD bands in 
$^{175}$Hf, the bands in this nucleus are interpreted in the CRMF framework as near-prolate. 
However, it was concluded that some SD bands in less neutron-rich Hf isotopes may be 
triaxial. 

  Theoretical uncertainties in the description of the physical observables of rotational 
bands (deformations, moments of inertia and their evolution with spins, band crossing
frequencies and alignment gains at band crossings etc) and their dependence on CEDF have
been analyzed in Ref.\ \cite{A.14-jpg}.

\section{Conclusions}

The systematic studies of different types of rotational bands (normal-deformed,
smooth terminating, superdeformed and hyperdeformed) have been performed in 
one-dimensional cranking approximation of the CDFT in different regions of 
nuclear chart from ground states to the extremes of deformation and spin. 
Contrary to the studies of rotating nuclei in non-relativistic DFT models, the 
CDFT studies have been performed systematically covering large amount 
of experimental data in the region of study. The CRMF and CRHB+LN calculations 
succesfully describe the experimental situation and in many cases they outperform 
non-relativistic models. The rotating nuclei offer a unique 
laboratory for testing of the channels of the CDFT which are not accessible by
other physical observables. This is because their properties sensitively depend
on nuclear magnetism (time-odd mean fields) and underlying single-particle
structure. The part of the success of the CDFT in the description of rotating 
nuclei is  definitely attributable to the fact that time-odd mean fields are 
uniquely defined via Lorentz covariance.

\vspace{1.0cm}

\centerline{\bf Acknowledgements}
\vspace{0.6cm}

This material is based upon work supported
by the U.S. Department of Energy,
Office of Science, Office of Nuclear Physics under
Award No.\ DE-SC0013037.

%
%

\bibliography{afanasjev}

\end{document}